\documentclass{mn2e}
\usepackage{psfig}
\usepackage{longtable}

\def\ltsima{$\; \buildrel < \over \sim \;$}
\def\lsim{\lower.5ex\hbox{\ltsima}}
\def\gtsima{$\; \buildrel > \over \sim \;$}
\def\gsim{\lower.5ex\hbox{\gtsima}}
\def\mes{M\'esz\'aros}

\def\bsax{{\it Beppo}SAX}
\def\batse{BATSE}
\def\bb{black--body}

\def\ama{$E_{\rm peak}-E_{\rm iso}$}
\def\ghi{$E_{\rm peak}-E_{\gamma}$}
\def\pl{power--law}
\def\syn{synchrotron}

\begin{document}

\title[Black--bodies in GRB prompt spectra]
{Black--body components in Gamma--Ray Bursts spectra?}

\author[Ghirlanda et al.]  {G. Ghirlanda$^1$\thanks{E--mail:
    giancarlo.ghirlanda@brera.inaf.it}, Z. Bosnjak$^{2,1}$,
  G. Ghisellini$^1$, F. Tavecchio$^1$ and C. Firmani$^{1,3}$\\
  $^{1}$Osservatorio Astronomico di Brera, via E. Bianchi 46, I--23807 Merate, Italy \\
  $^2$ Institut d'Astrophysique de Paris, 98 bis bd Arago, 75014 Paris, France  \\
  $^{3}$Instituto de Astronom\'ia, U.N.A.M., A.P. 70-264, 04510, M\'exico,
  D.F., M\'exico} \maketitle

\begin{abstract}
  We study 7 Gamma Ray Bursts (GRBs), detected both by the BATSE instrument,
  on--board the Compton Gamma Ray Observatory, and by the Wide Field Camera
  (WFC), on board $Beppo$SAX.  These bursts have measured spectroscopic redshifts
  and are a sizeable fraction of the bursts defining the correlation between
  the peak energy $E_{\rm peak}$ (i.e. the peak of the $\nu F_\nu$ spectrum)
  and the total prompt isotropic energy $E_{\rm iso}$ (so called ``Amati''
  relation).  Recent theoretical interpretations of this correlation assume
  that black--body emission dominates the time resolved spectra of GRBs, 
  even if, in
  the time integrated spectrum, its presence may be hidden by the change of
  its temperature and by the dilution of a possible non--thermal power law
  component.  We perform a time resolved spectral analysis, and show that
  the sum of a power--law and a black--body gives acceptable fits to the time
  dependent spectra within the BATSE energy range but overpredicts the
  flux in the  WFC X--ray range. Moreover, a fit with a cutoff power--law plus
  a black--body is consistent with the WFC data but the black--body 
  component contributes a negligible fraction of the total flux. 
  On the contrary, we find that fitting the spectra with a Band model or a
  simple cutoff power--law model yields an X--ray flux and spectral slope which
  well matches the WFC spectra.
\end{abstract}


\section{Introduction}

Our knowledge of the spectral properties of the prompt emission of long Gamma
Ray Bursts (GRBs) is based on systematic studies of their time integrated
spectrum (e.g. Golenetskii et al., 1983; Band et al. 1993; Barraud et al.
2003; Amati et al. 2002) which turned out to be described by two smoothly
joint \pl s with typical photon indices $\alpha\simeq -1$ and $\beta\simeq -2.5$
for the low and high energy components, respectively, or a cutoff \pl.
Several studies aiming to characterise the time dependent behaviour of the
spectrum (Ford et al. 1995; Preece et al. 1997; Crider et al. 1997; Ghirlanda,
Celotti \& Ghisellini 2002, Ryde et al. 2000; Kaneko et al. 2006) have
demonstrated that the overall spectral shape and the peak energy $E_{\rm
  peak}$ (in a $\nu F(\nu)$ plot) evolve in time.  The evolution is rather
complex and there is no unique trend, but a prevalence of a hard--to--soft
behaviour is observed.

The time integrated properties, however, are the one used to calculate the
bolometric emitted energy of GRBs (both isotropic, $E_{\rm iso}$, and collimation
corrected, $E_\gamma$) and to relate them to the peak energy $E_{\rm peak}$
(the so called ``Amati", ``Ghirlanda" and ``Liang \& Zhang" relations -- Amati
et al. 2002; Ghirlanda, Ghisellini \& Lazzati 2004,
hereafter GGL04; Ghirlanda et al. 2007; Liang \& Zhang 2005).
Furthermore, even the correlation between the isotropic peak luminosity
$L_{\rm peak}$ and $E_{\rm peak}$ (the so--called ``Yonetoku" relation --
Yonetoku et al. 2004), and the $L_{\rm peak}$--$E_{\rm peak}$--$T_{0.45}$
relation (the so--called ``Firmani" relation -- Firmani et al. 2006) make use
of the time integrated spectrum (see Ghirlanda et al.  2005).  The fact that
these correlations apply to the time integrated spectrum, even if it evolves
in time, may underline some global property of the burst.

In this respect there have been, very recently, important suggestions and new
ideas for explaining the ``Amati", the ``Ghirlanda" and also the ``Firmani"
relation. The simplest way to have a relation between the emitted energy or
luminosity and $E_{\rm peak}$ is through \bb\ emission.  Indeed, in this case,
the number of free parameters is kept to a minimum: the rest frame bolometric
and isotropic \bb\ luminosity would depend on the emitting surface, the
temperature and the bulk Lorentz factor.  Any other emission process would
depend on some extra parameters, such as the magnetic field and/or the
particle density, and it would then be more difficult, if these quantities
vary from burst to burst, to produce a correlation with a relatively small
scatter such as the \ghi\ one.

Rees \& \mes\ (2005), Thompson (2006) and Thompson, \mes\ \& Rees
(2006) explain these correlations assuming that a considerable
fraction of the prompt emission flux is due to a black--body.  This
does not imply, however, that the entire observed flux is a single
black--body (we already know that this is not the case).  

Indeed, time integrated GRB spectra are typically described by the
Band model or cutoff--power law model.  The time integrated spectrum,
however, being the result of the spectral evolution, could be best
fitted by a model which is not the same used for the time resolved
spectra. Within the black--body interpretation, there could be at least
two alternatives: the time integrated spectrum (which looks like a
cutoff--power law or a Band model) is (a) the result of the
superposition of different black--bodies with a time dependent
temperature and flux or (b) the sum of two components, i.e. one
thermal (black--body) and one non--thermal (power law or double
power law) as suggested by Ryde 2004. In both cases, since the
temperature of the single (time resolved) black--bodies and/or the
slope of the \pl\ can evolve in time, the time--integrated spectrum
could well be modelled by a smoothly broken \pl\ (i.e. the Band
function, see below), hiding the presence of the black--body.  This
requires to perform the time resolved spectral analysis in order to
assess the presence of an evolving black--body component possibly with
a non--thermal \pl\ component.

Evidences of the presence of a thermal \bb\ component were discovered
in the BATSE spectra (e.g. Ghirlanda, Celotti, Ghisellini 2003,
hereafter GCG03).  This component dominates the initial emission phase
up to $\sim 2$ s after the trigger.  During this phase the luminosity
and the temperature evolve similarly in different GRBs while the late
time spectrum is dominated by a non thermal component (e.g. it is
fitted with the empirical Band et al. 1993 model).  Attempts to
deconvolve these spectra with a mixed model, i.e. a \bb\ plus a \pl\
(Ryde et al.  2005), showed that the \bb\ (albeit with a monotonically
decreasing flux) could be present also during later phases of the
prompt emission (see also Bosnjak et al. 2005).

As a test of the recently proposed ``black--body'' interpretations of
the \ama\ and \ghi\ correlations, we consider, among the sample of
GRBs used to define these correlations, those bursts that were
detected by BATSE and with published WFC spectra.  Given the
relatively large brightness of these bursts, it is possible for them
to meaningfully analyse the time dependent properties of their
spectra.

The focus of this paper is not much on the study of the spectral
evolution of these few bursts\footnote{ The analysis of how the
spectral parameters evolve in time with respect to the \ama\ and \ghi\
correlations is the content of a forthcoming paper (Bosnjak et al., in prep.).},
but, instead, on the relevance of the \bb\ in the time resolved
spectra together with the relevance of the sum of the black--bodies,
possibly at different temperatures, in the time integrated spectrum.
To this aim we adopt for our analysis a \pl+\bb\ model, besides the
``standard'' Band and cutoff \pl\ model.  We anticipate that the
\pl+\bb\ model, although giving acceptable fits, is inconsistent with
the WFC data.  A more complex fit, made by adopting the sum of a \bb\
and a cutoff \pl, is equally acceptable and consistent with the WFC
data, but implies that the \bb\ flux is a minor fraction of the total.

The paper is organised as follows: in \S 2 we recall the basic ideas of the
``black--body'' interpretation of the \ama\ and \ghi\ correlations; in \S 3 we
introduce the sample and the spectral analysis procedure; in \S 3 we present
the results of the time resolved spectral analysis and the comparison of the
BATSE and WFC spectra with the three adopted model.  We discuss our results in
\S 4.

\section{The interpretation of the spectral--energy correlations}

The recent theoretical attempts to explain the spectral--energy relations, and
in particular the \ama\ (Amati) one, largely motivate the present work.
Therefore it may be useful to summarise here the arguments put forward by
Thompson (2006) and by Thompson, \mes\ \& Rees (2006).

Consider a fireball that at some distance $R_0$ from the central engine is
moving relativistically with bulk Lorentz factor $\Gamma_0$.  As an example,
one can think that $R_0$ is the radius of the progenitor star.  Assume that a
large fraction of the energy that the fireball dissipates at $R_0$ is
thermalized and forms a \bb\ of luminosity:
\begin{equation}
L_{\rm BB, iso}\, =\, \pi R_0^2 \Gamma_0^2 \sigma {T'_0}^4\, = \,
\pi {R_0^2\over \Gamma_0^2} \sigma T_0^4 
\end{equation}
where $T'_0$ and $T_0=\Gamma_0 T'_0$ are the temperatures at $R_0$ in
the comoving and observing frame, respectively.  The collimation
corrected luminosity is $L_{\rm BB} =(1-\cos\theta_{\rm j})L_{\rm BB,
iso}$, which, for small semiaperture angles of the jetted fireball
(assumed to be conical) gives
\begin{equation}
\theta_{\rm j}^2 \, \sim \, { 2 L_{\rm BB}\over L_{\rm BB, iso}}
\end{equation}
Now Thompson (2006) and Thompson, \mes~ \& Rees (2006) introduce one
key assumption: for causality reasons 
$\Gamma_0 \sim 1/\theta_{\rm j}$.  
This allows to substitute $\Gamma_0$ in Eq. (1) to obtain:
\begin{equation}
L_{\rm BB, iso}\, \sim \, 
2 \pi R_0^2 { L_{\rm BB}\over L_{\rm BB, iso}  } \sigma {T_0}^4 
\end{equation}
Setting $E_{\rm BB, iso}=L_{\rm BB, iso} t_{\rm burst}$ and 
$E_{\rm BB}=L_{\rm BB} t_{\rm burst}$, where $t_{\rm burst}$ is 
the duration of the prompt emission, one has
\begin{equation}
E_{\rm peak}\, \propto\, T_0\, \propto\,  
E_{\rm BB, iso}^{1/2}  E_{\rm BB}^{-1/4} t_{\rm burst}^{-1/4}
\end{equation}
This reproduces the ``Amati" relation {\rm if} $E_{\rm BB}$ is nearly
the same in different bursts and if the dispersion of the GRB duration
is not large.  One can see that a key assumption for this derivation
is the black--body law.  It is the $L\propto T^4$ relation which
allows to derive $E_{\rm peak} \propto E_{\rm iso}^{1/2}$.

\section{Sample selection and analysis}

We consider all bursts detected by BATSE with measured spectroscopic redshift
which were also detected by \bsax\ and for which the WFC data were published
(Amati et al. 2002; Frontera, Amati \& Costa 2000).  In Tab. 1 we list our
bursts and their time integrated spectral properties as found in the
literature. We also report the duration ($T_{90}$) derived from the BATSE
$\gamma$--ray light curve, the 50--300 keV energy fluence and the hard X--ray
(2--28 keV) energy fluence. We include in our sample also GRB 980329 and
980326 for which only a range of possible redshifts (the most accurate for
980326) were found (see also GGL04). 

\begin{table*} 
\begin{center}
\begin{tabular}{lllllllll}
\hline
GRB   &$z$      &$\alpha$  &$\beta$ &$E_{\rm peak}$   &REF            &$T_{90}$        &$F$(50--300keV)     &$F$(2--28keV)      \\
      &         &          &        &      keV        &               &    s           &erg cm$^{-2}$       &erg cm$^{-2}$    \\
\hline
970508 &0.835    &$-$1.71 (0.1)   &$-$2.2 (0.25)  &79 (23)      &1, 8   &              &                    &                    \\
       &0.835    &$-$1.19         &$-$1.83        &$>1800$      &1, 9   &23.1$\pm$3.8  &1.1$\times 10^{-6}$ &8.3$\times 10^{-7}$ \\
971214 &3.418    &$-$0.76 (0.1)   &$-$2.7 (1.1)   &155 (30)     &2, 8   &31.23$\pm$1.18&6.44$\times 10^{-6}$&3.2$\times 10^{-7}$ \\
980326 &0.9--1.1 &$-$1.23 (0.21)  &$-$2.48 (0.31) &33.8 (17)    &3, 8   &...           &...                 &5.5$\times 10^{-7}$  \\
980329 &2--3.9   &$-$0.64 (0.14)  &$-$2.2 (0.8)   &233.7 (37.5) &4, 8   &18.56$\pm$0.26&3.2$\times10^{-5}$  &4.3$\times 10^{-6}$ \\
980425 &0.0085   &$-$1.26         &               &120          &5, 9   &34.88$\pm$3.78&2.47$\times10^{-6}$ &1.8$\times 10^{-6}$ \\
990123 &1.6      &$-$0.89 (0.08)  &$-$2.45 (0.97) &781 (62)     &6, 8   &63.4$\pm$0.4  &1.0 $\times10^{-4}$ &9.0$\times 10^{-6}$ \\
990510 &1.602    &$-$1.23 (0.05)  &$-$2.7 (0.4)   &163 (16)     &7, 8   &68$\pm$0.2    &1.1$\times10^{-5}$  &5.5$\times 10^{-6}$ \\
\hline
\hline
\end{tabular}                                                                                           
\caption{ Time integrated properties of the bursts with spectroscopic
  redshift and detected by both \batse\ and \bsax\ and with published
  \bsax--WFC spectra.  The duration $T_{90}$ and the (50--300keV)
  fluence [$F$(50--300keV)] are from the on--line BATSE catalogue. The
  2--28 keV fluence is reported from Tab. 1 of Amati et al. 2002 for
  all bursts except GRB 980425 for which we report the 2--26 keV
  fluence given in Pian et al. 2000. In the case of GRB~980326 we
  could not find these information in the publicly available
  archive. For GRB~970508 we report the spectral results of the \bsax\
  data (first line) and the results obtained from the BATSE data
  (second line). First set of references is for the redshift: 1)
  Metzger et al.  1997; 2) Kulkarni et al. 1998; 3) Bloom et al. 1999;
  4) Lamb et al. 1999 (and references therein); 5) Galama et al.
  1998; 6) Kulkarni et al. 1999; 7) Vreeswijk et al. 2001; Second set
  of references is for the spectral parameters: 8) Amati et al. 2002;
  9) Jimenez, Band \& Piran 2001.  }
\end{center}
\label{tab1}
\end{table*}                                                                                                            

For all the bursts we analysed the BATSE Large Area Detector (LAD) spectral
data which consist of several spectra accumulated in consecutive time bins
before, during and after the burst.  Only for GRB~990123 we analysed the
Spectroscopic Detectors (SD) data because of a gap in the LAD data sequence.
The spectral analysis has been performed with the software \emph{SOAR} v3.0
(Spectroscopic Oriented Analysis Routines by Ford et al. 1993), which we
implemented for our purposes.

For each burst we analysed the \batse\ spectrum accumulated over its total
duration (which in most cases corresponds to the $T_{90}$ parameter reported
in the BATSE catalogue) and the time resolved spectra distributed within this
time interval.  The time resolved spectra are accumulated on--board according
to a minimum signal--to--noise criterion with a minimum integration time of
128 ms.  As the bursts of our sample have quite large fluence (i.e. $\ge
10^{-6}$ erg cm$^{-2}$ integrated over the 50--300 keV range) in most cases we
could analyse their time resolved spectra as they were accumulated by the
on--board algorithm. Only the spectra at the beginning or at the end of the
bursts (or during interpulses phases) were accumulated in time in order to
have a larger signal.  Energy rebinning (i.e. at least 30 (15) counts per bin
for the LAD (SD) spectra) was systematically applied in our analysis in order
to test the goodness of the fits through the $\chi^2$ statistics.

The adopted spectral analysis procedure is the standard forward--folding which
folds the model spectrum with the detector response and, by varying the model
free parameters, minimises the $\chi^2$ between the model and the data.  This
procedure requires the knowledge of the background spectrum corresponding to
each analysed spectrum. In order to account for the possible time variability
of the background we modelled it as a function of time.  We selected two time
intervals (before and after the burst) as close as possible to the burst (not
contaminated by the burst itself) of typical duration $1000$ s.  We fit the
spectra contained in these intervals with a background model which is a
polynomial function of time $B(E,t)$, and, being a spectrum, also of the
energy $E$.  Each energy bin of the spectra selected for the background
calculation is interpolated with this polynomial function.  This fit is tested
for by inspecting the distribution of its $\chi^2$ as a function of energy.
In this way we obtain the best fit time--dependent background model function
$B_{best}(E,t)$ which is extrapolated to the time interval $\Delta t$ of each
time resolved spectrum and subtracted to the data. This method is the same
adopted in previous analysis of the BATSE data (e.g. Preece et al. 2000;
Kaneko et al. 2006).

\subsection{Spectral models}

For the analysis of both the time resolved and the time integrated spectra we
use three models which were already tested in fitting the BATSE spectral data
(Preece et al. 2000; Ghirlanda et al. 2002; Ryde 2004; Kaneko 2006):
\begin{enumerate}
\item The Band (B) model (originally proposed by Band et al. 1993) which
consists of 2 power laws joined smoothly by an exponential roll--over.
Its analytical form is:
\begin{eqnarray}
N(E) & = & A E^\alpha \exp \left( - {E \over E_0} \right); \quad
     {\rm for} \  E \leq \left( \alpha - \beta \right) E_{0}    \nonumber \\ 
N(E) & = & A E^{\beta} [(\alpha-\beta)E_0]^{\alpha-\beta} \exp(\beta-\alpha);  \nonumber \\ 
     &   & {\rm for} \;\; E \geq \left( \alpha - \beta \right) E_{0}  
\end{eqnarray}
%
The free parameters,
which are the result of the fit to the observed spectrum, are: the
normalisation of the spectrum $A$;
the low energy power law photon spectral index $\alpha$; the high
energy power law photon spectral index $\beta$ and the break energy, which
represents the e--folding parameter, $E_{0}$. 
If $\beta<-2$ and $\alpha > -2$ this model has a peak in the 
$EF_E$ representation which is $E_{\rm peak}=(\alpha +2)E_{0}$.  
In the fits we assume that $\alpha$ and $\beta$ can vary within the range 
[$-5$, 1] while the break energy is allowed to vary in the same range 
covered by the spectral data, i.e. $\sim$ 30--1800 keV.  
The B model is a fair representation of the spectral model produced in
the case of emission from a relativistic population of electrons, distributed
in energy as a single or a broken power law, emitting synchrotron and/or
inverse Compton radiation, and can also reproduce the case of an electron
energy distribution which is a Maxwellian at low energies and a power law at
high energies, emitting \syn\ radiation (e.g. Tavani et al. 1996).

\item 
The cut--off power law (CPL) is composed by a power law
ending--up in an exponential cutoff.  It corresponds to the previous
Band model without the high energy \pl\ component.  Its form is:
\begin{equation}
     N(E)=A E^{\alpha} \exp\left( -\frac{E}{E_{0}}\right)
\end{equation}
The free parameters are the same of the Band model without the high
energy component. 
If $\alpha >-2$ also this model presents a peak in its $EF_E$ 
representation which is $E_{\rm peak}=(\alpha + 2)E_{0}$. 
This model can represent the case of thermal or quasi--thermal
Comptonization, even when saturated (i.e. a Wien spectrum,
with $\alpha=2$).

\item 
The black--body + \pl\ (BBPL) model is
\begin{equation}
N(E)=A \frac{E^2}{\exp(E/kT)-1} + B E^{\alpha}
\end{equation} 
where $\alpha$ is the spectral index of the \pl; $kT$ the \bb\ temperature
and $A$ and $B$ are the normalisations of the two spectral components.
In this case, the peak of the $\nu F_{\nu}$ spectrum depends on the relative
strength of the two model components and on the spectral energy range where
the spectrum is considered.  The peak energy of the \bb\ component only is
$E_{\rm peak, BB}=3.93 kT$ (in $\nu F_\nu$).  The (simplest) physical
rationale of this model is the possible different origin of the two
components: the thermal \bb\ emission could be photospheric emission from the
fireball (e.g. Daigne \& Mockovitch 2000) while the \pl\ component might be
the non--thermal emission from relativistic electrons accelerated above the
photosphere at the classical internal shock radius (see also Pe'er, Meszaros
\& Rees 2006).  The BBPL model is the simplest spectral model which combines a
thermal and a non--thermal component.  In \S 5 we will also discuss the more
complex case of a cutoff \pl\ plus a \bb\ component.

\end{enumerate}

Note that the number of free parameters is the same (i.e. four, including
normalisations) in the B and BBPL model while the CPL model has one less free
parameter.

The \batse\ spectra were fitted in the past with all these models.  Band et
al. (1993) proposed the B function to fit the time integrated spectra of
bright long GRBs.  Also the time resolved spectra could be fitted by either
the B or the CPL model (Ford et al. 1995; Ghirlanda, Celotti \& Ghisellini
2002).  More recently Kaneko et al. (2006) performed a systematic analysis of
the time resolved spectra of a large sample of BATSE bursts selected according
to their peak flux and fluence. From these works it results that the typical
low energy spectral slope (in both the B and CPL model) has a wide
distribution centred around $\alpha\sim -1$ with no preference for any
specific value predicted by the proposed emission models (i.e.  $\alpha=-2/3$
for optically thin synchrotron -- Katz et al. 1994; $\alpha=-3/2$ for
synchrotron cooling -- Ghisellini \& Celotti 1999; $\alpha=0$ for jitter
radiation -- Medvedev 2000).  The high energy photon spectral index $\beta$
has a similar dispersion (i.e. 0.25) of the $\alpha$ distribution and its
typical value is --2.3.  The peak energy has a narrow ($\sigma\le 100$ keV)
distribution centred at $\sim$ 300 keV.  A small fraction (7\%) of the time
resolved spectra have $\beta>-2$ which means that the peak energy of the
$EF_E$ spectrum is above the upper energy threshold (i.e. $E_{\rm peak}>2$
MeV).  The composite BBPL model was fitted to the time resolved spectra of a
few bright \batse\ bursts (Ryde 2005, Bosnjak et al. 2005).

In the following section we present the spectral parameters of the fits
obtained with the three models above. The scope of this paper is not to decide
which (if any) of the proposed models best fits the spectra.  It has been
already shown (e.g.  Ryde 2005) that the time resolved BATSE spectra can be
adequately fitted with both the B(CPL) model and the BBPL model.

\section{Results}

We here show the spectral evolution and compare the spectral parameters of
the three models described in \S 3. We also compare the spectral results of
our analysis of the BATSE time integrated spectra (reported in Tab. 2) with the
results gathered from the literature (Tab. 1). We then discuss the
contribution of the \bb\ component to the spectrum and compare the spectral
fits of the three models with the constraints given by the WFC data.

\subsection{Spectral evolution}

We present the spectral evolution of the fit parameters obtained with
the three models described in \S 3. 

\subsubsection{GRB 970508}

The spectral parameters of the time integrated spectrum published in Amati et al.
(2002) for GRB~970508 were found by the analysis of the WFC [2--28 keV] and
Gamma Ray Burst Monitor [GRBM, 40--700 keV] data and they differ from those
found by the BATSE spectral analysis and published in Jimenez et al. (2000).
We report the different results in Tab. 1.  The main difference is that
according to the \bsax\ spectrum this burst has a considerably low peak energy
while the \batse\ spectrum indicates that $E_{\rm peak}>$1800 keV. We have
re--analysed the \batse\ spectrum confirming the results found by Jimenez et
al. (2000).  In particular we found an unconstrained peak energy when fitting
both the B and CPL model.  The spectrum in the 40--700 keV energy range of
GRB~970508 presented in Amati et al. (2002) is composed by only two data
points with a quite large associated uncertainty.  In this case the fit (with
the B model) is dominated by the WFC spectrum, which does not present any
evidence of a peak (in $\nu F_\nu$) within its energy range.  Combining the
GRBM and WFC data Amati et al. (2002) found $E_{\rm peak}=79$ keV, but the
GRBM spectral data appear consistent also with an high energy component with
$\beta\ge -2$ (which is what is found from the fit of the BATSE spectrum).  If
the real GRB spectrum is that observed by BATSE this burst would be an outlier
for the Amati correlation (see also Fig. 3 of GGL04).  Given the possible
uncertainties of the \bsax\ spectrum, we do not consider this burst in the
following analysis because the BATSE spectrum does not allow to constrain its
peak energy.

\subsubsection{GRB 971214} 
\begin{figure}
\begin{center}
\psfig{figure=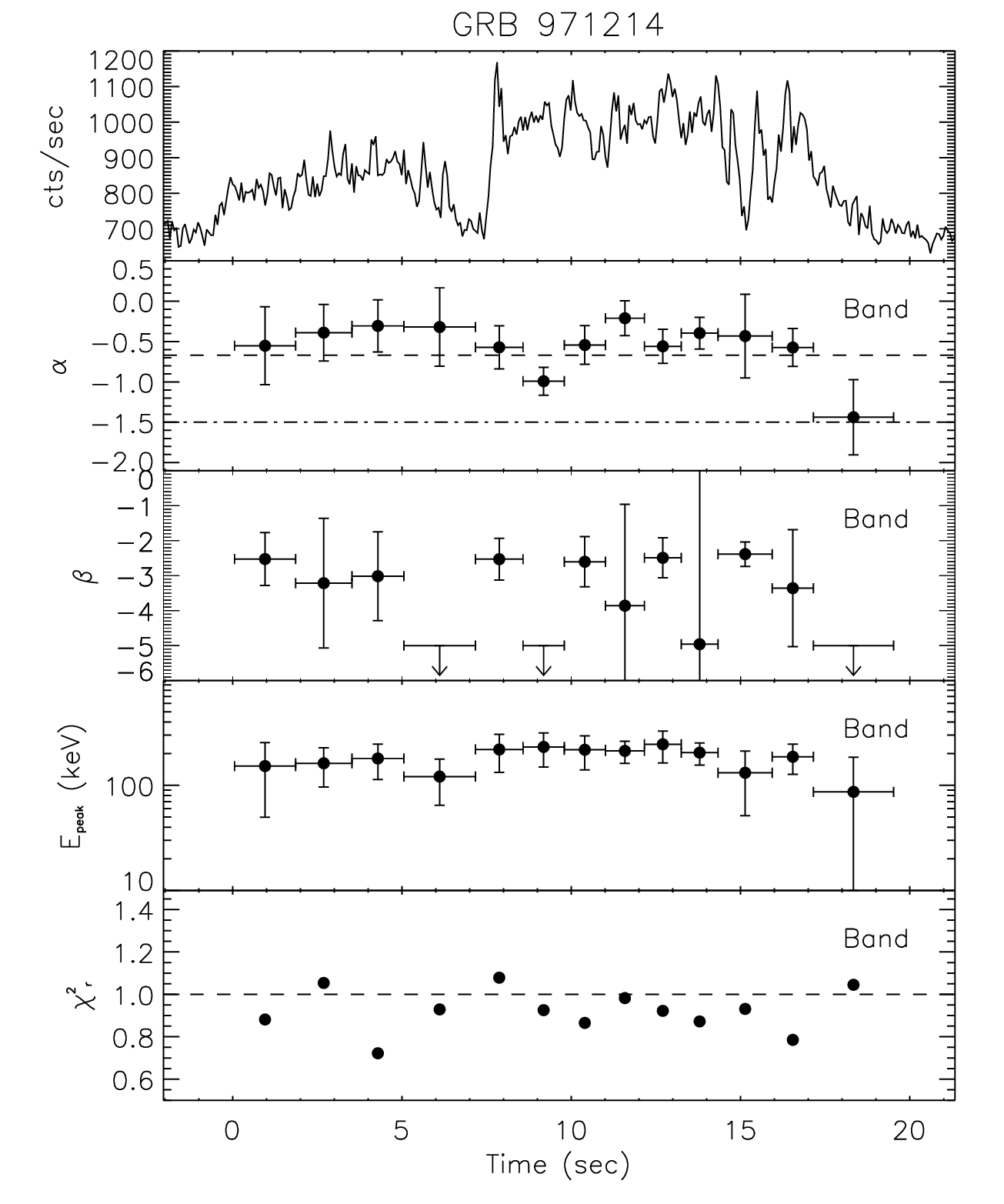,height=7.5cm,width=8.5cm}
\psfig{figure=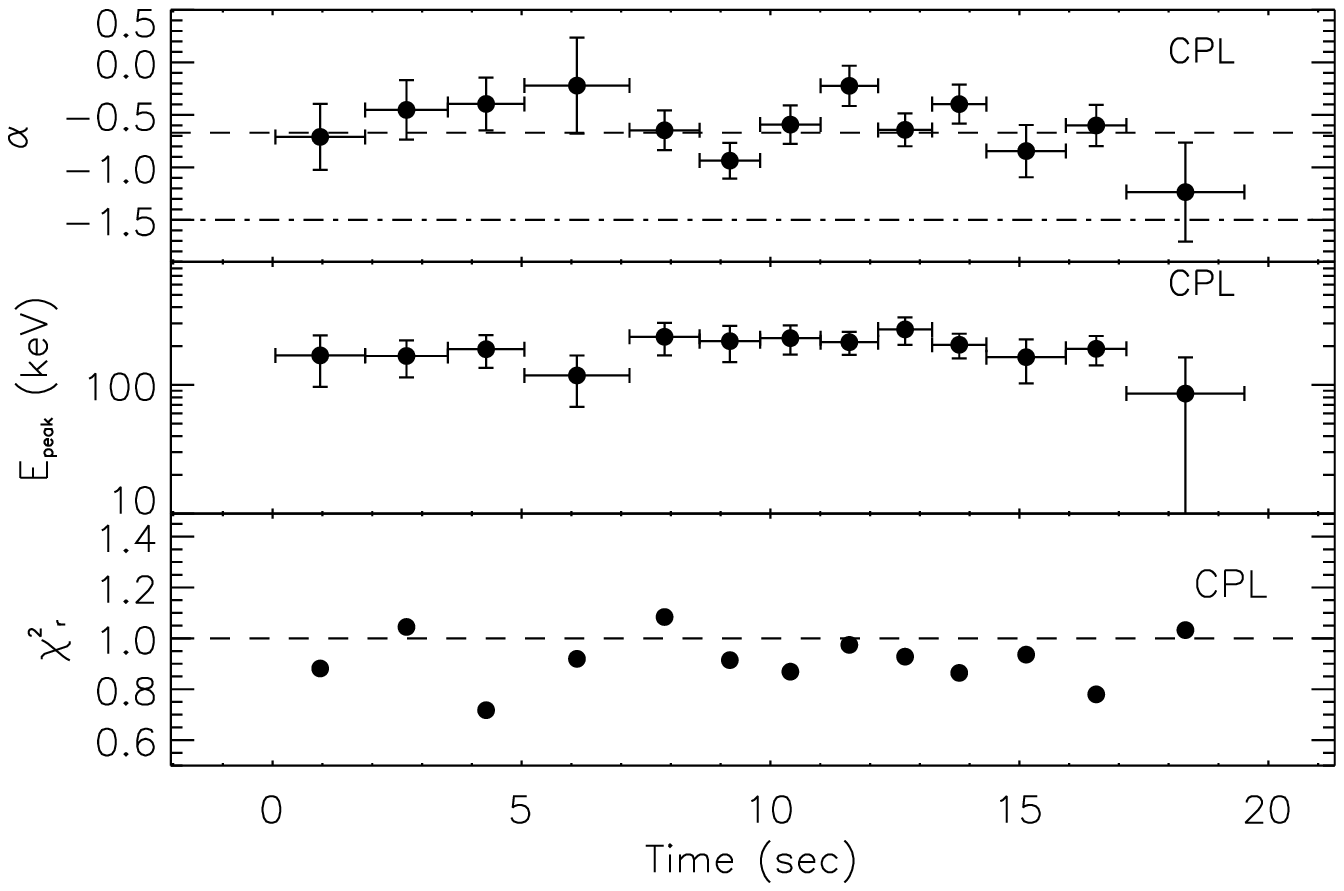,height=5cm,width=8.5cm}
\psfig{figure=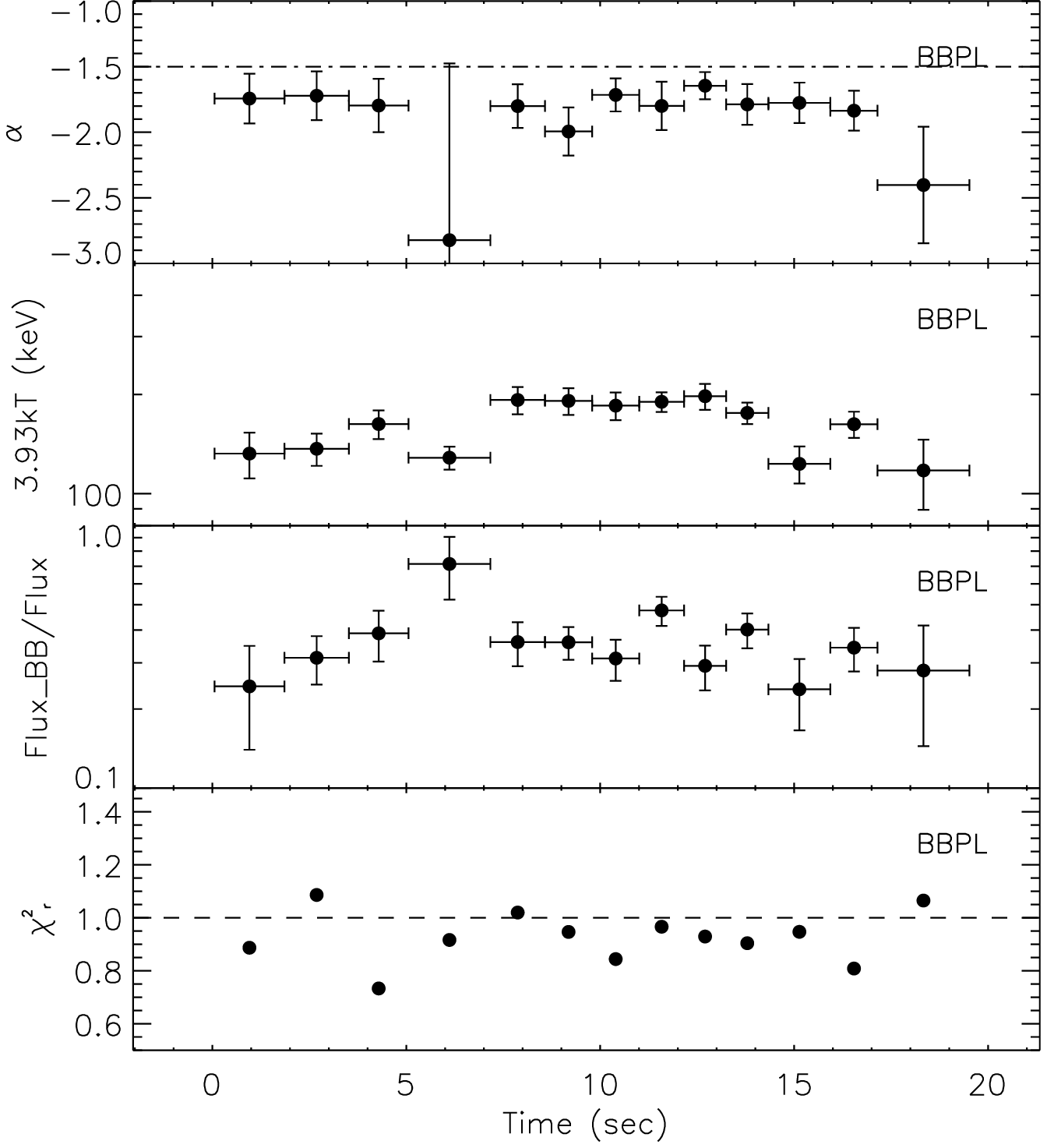,height=6.5cm,width=8.5cm}
\caption{
  Spectral evolution of GRB 971214. Top panels: Band model fit results. The
  first panel represents the light curve (in counts/s detected at energies
  $\ge$ 25 keV -- without the background subtraction). Mid panels:
  cutoff \pl\ fit results. Bottom panels: \bb+\pl\ fit results (we also
  report the contribution of the \bb\ component to the total flux in the
  observed 30 keV--2 MeV energy range). For all the three models we show for
  comparison the optically thin synchrotron limit ($\alpha=-2/3$,  dashed
  lines) and the case of synchrotron cooling ($\alpha=-3/2$, dot dashed
  line).  }
\label{971214evo}\end{center}
\end{figure}
\begin{figure}
\begin{center}
\psfig{figure=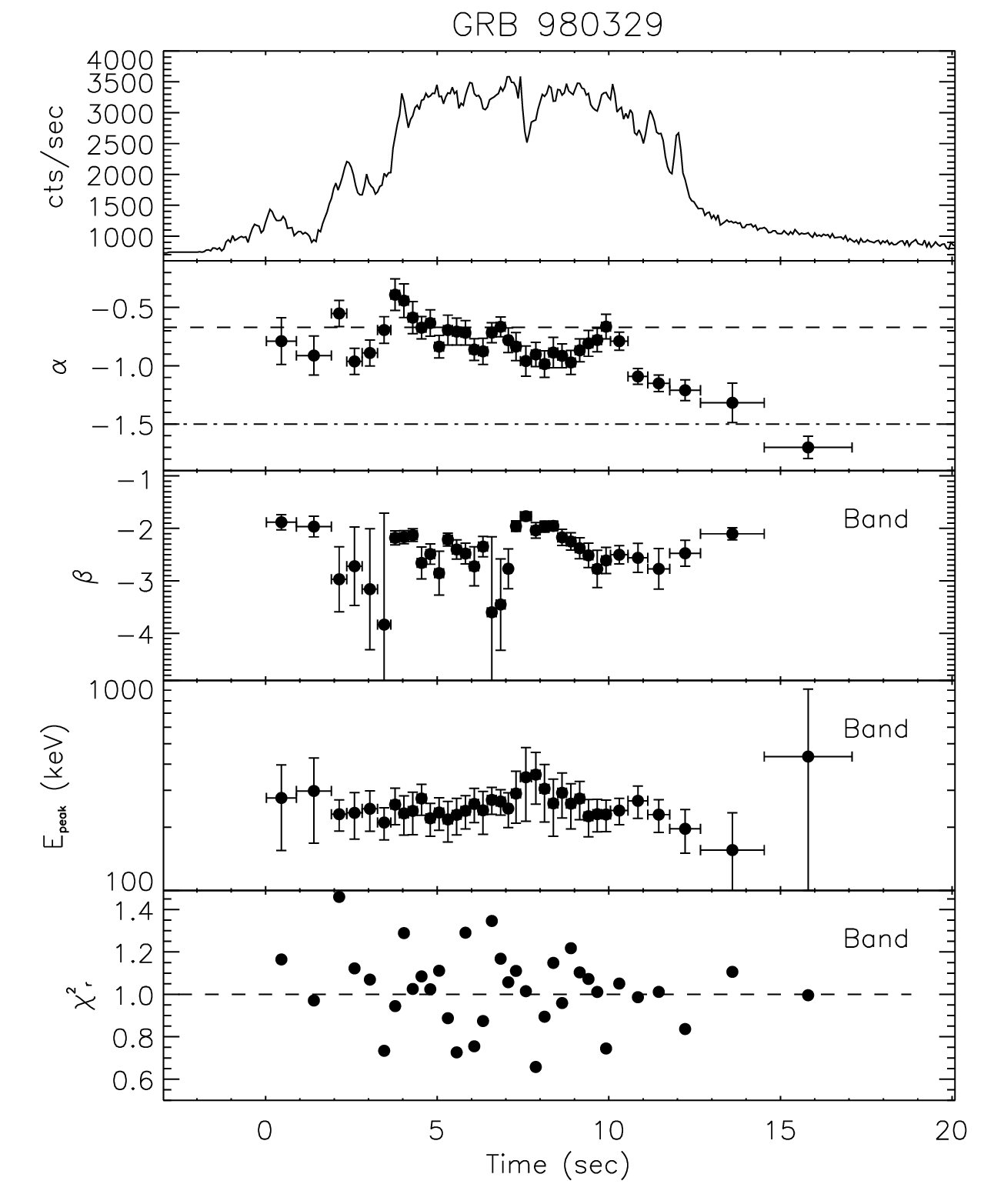,height=7.5cm,width=8.5cm}
\psfig{figure=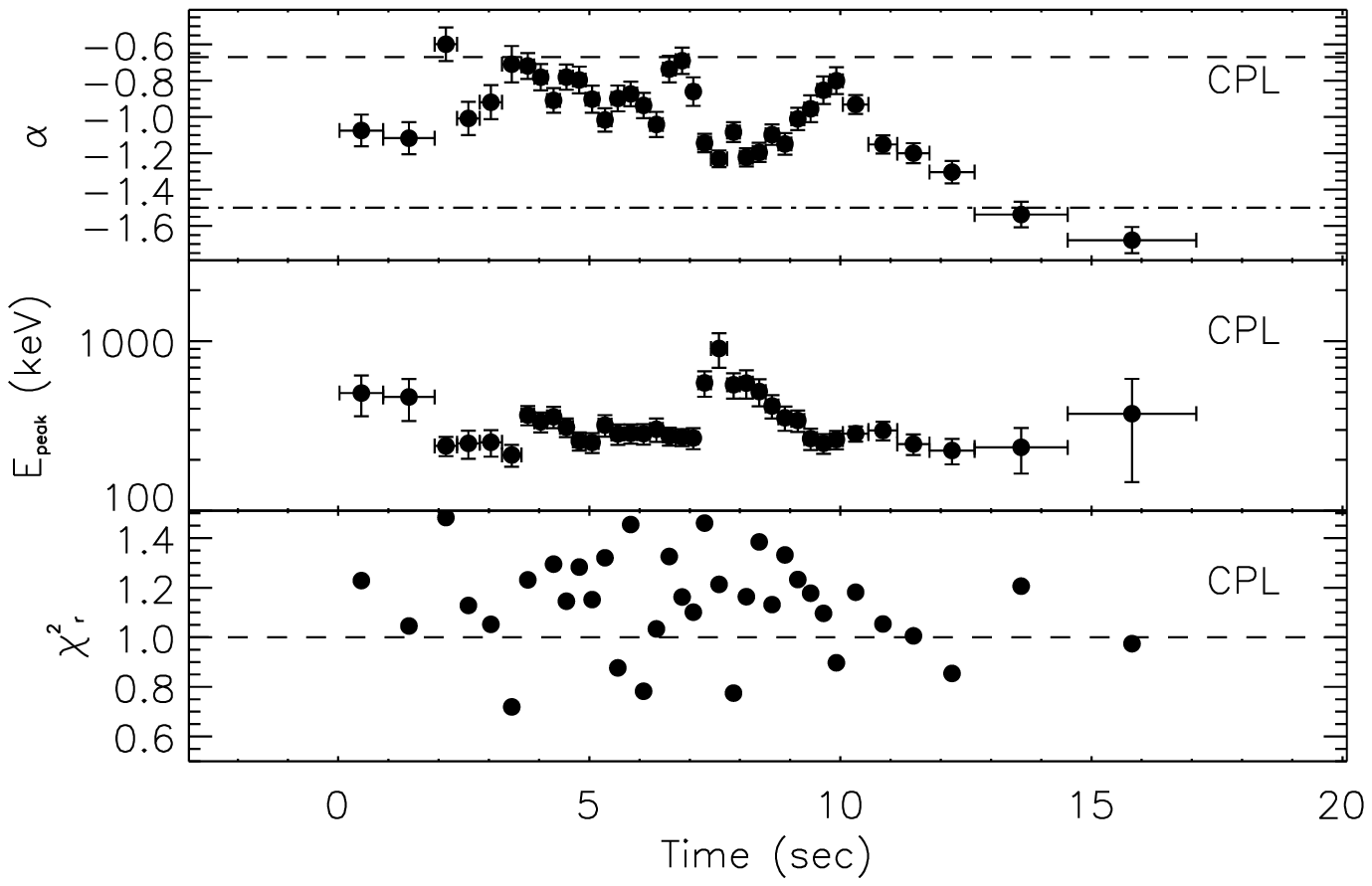,height=5cm,width=8.5cm}
\psfig{figure=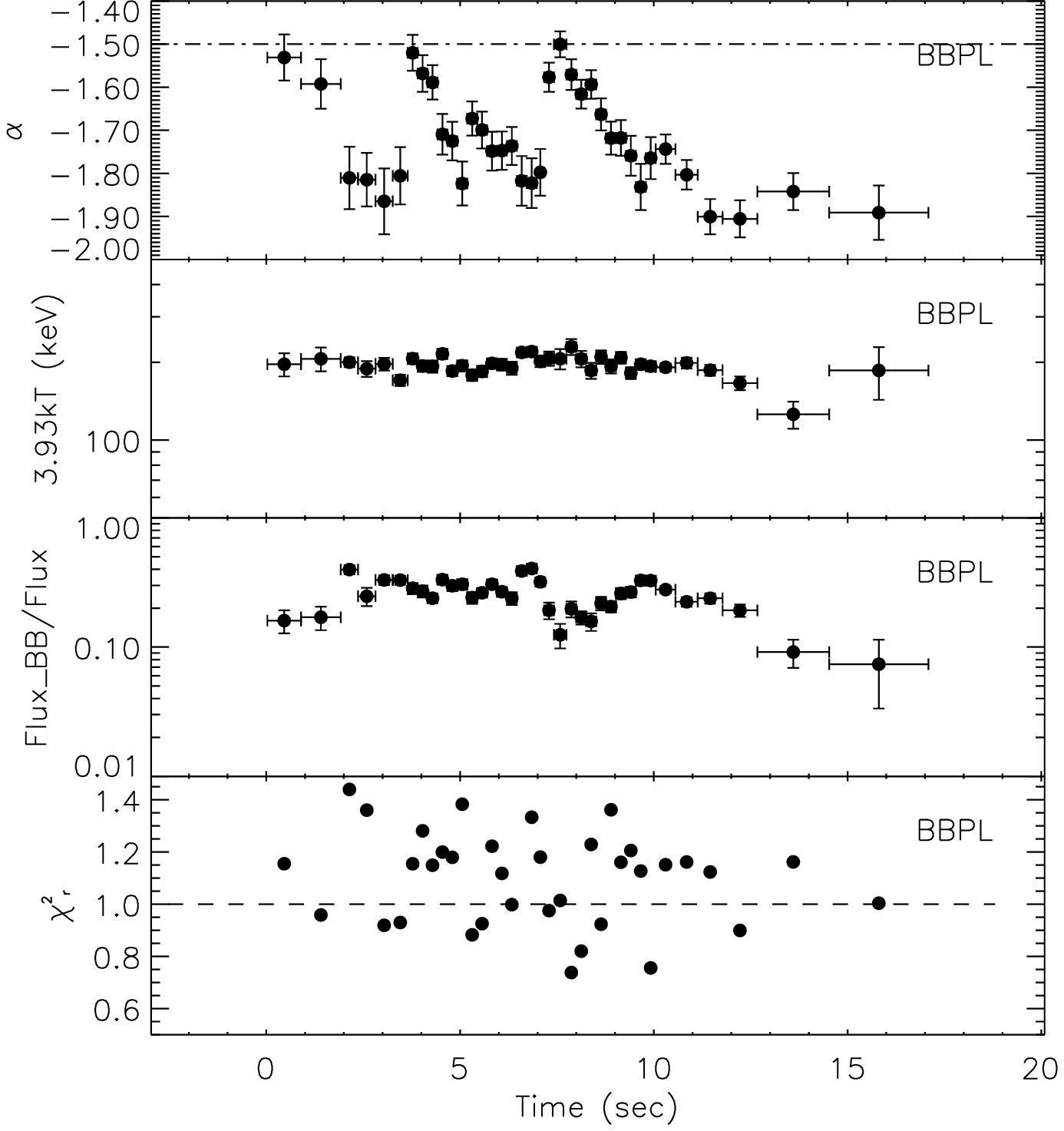,height=6.5cm,width=8.5cm}
\caption{
  Spectral evolution of GRB 980329. Symbols are the same of
  Fig.\ref{971214evo}.}
\label{980329evo}\end{center}
\end{figure}
\begin{figure}
\begin{center}
  \psfig{figure=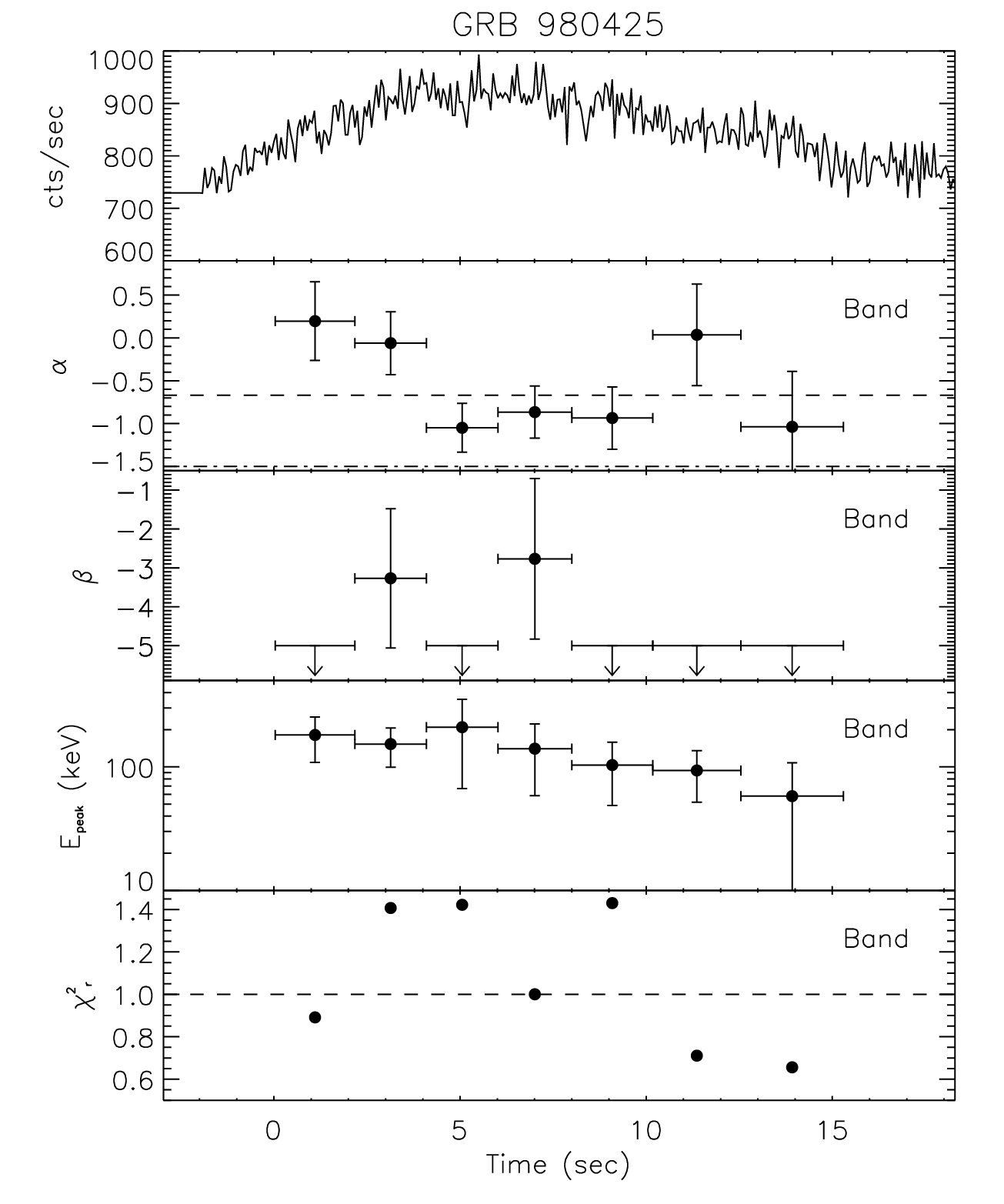,height=7.5cm,width=8.5cm}
\vskip -1 true cm
  \psfig{figure=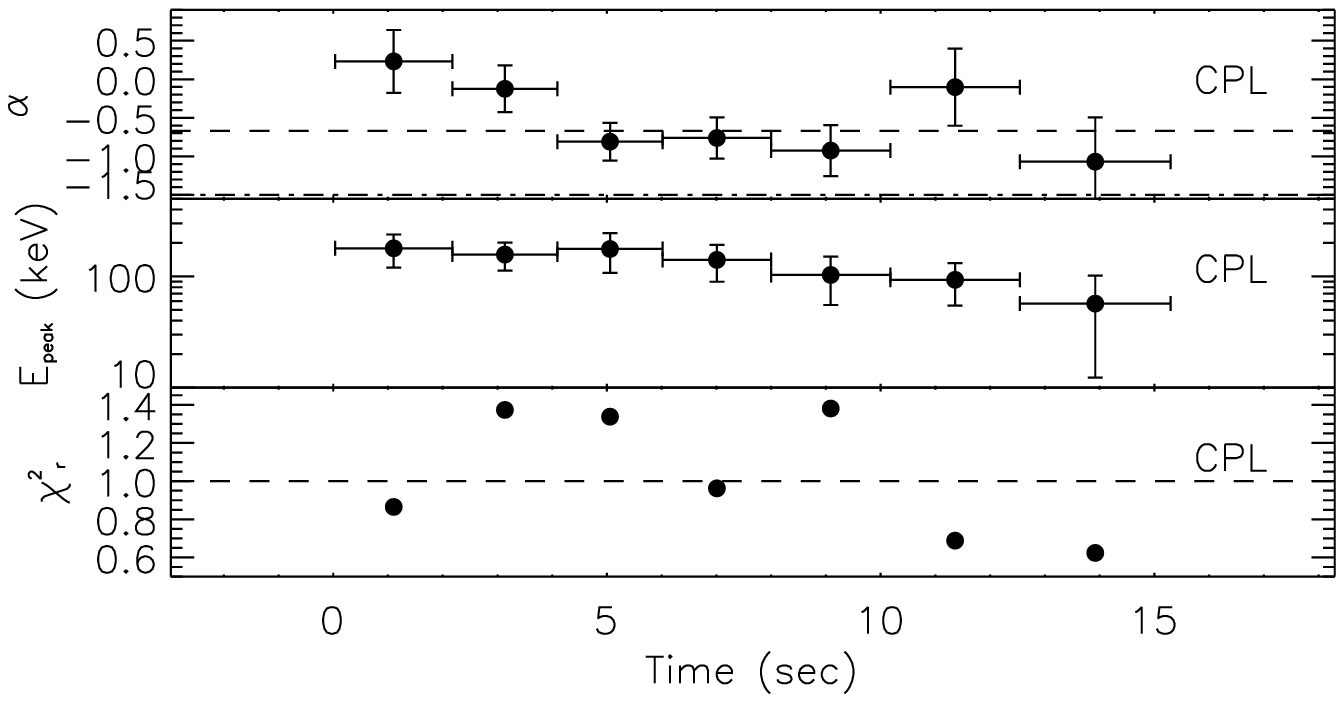,height=6cm,width=8.5cm}
  \psfig{figure=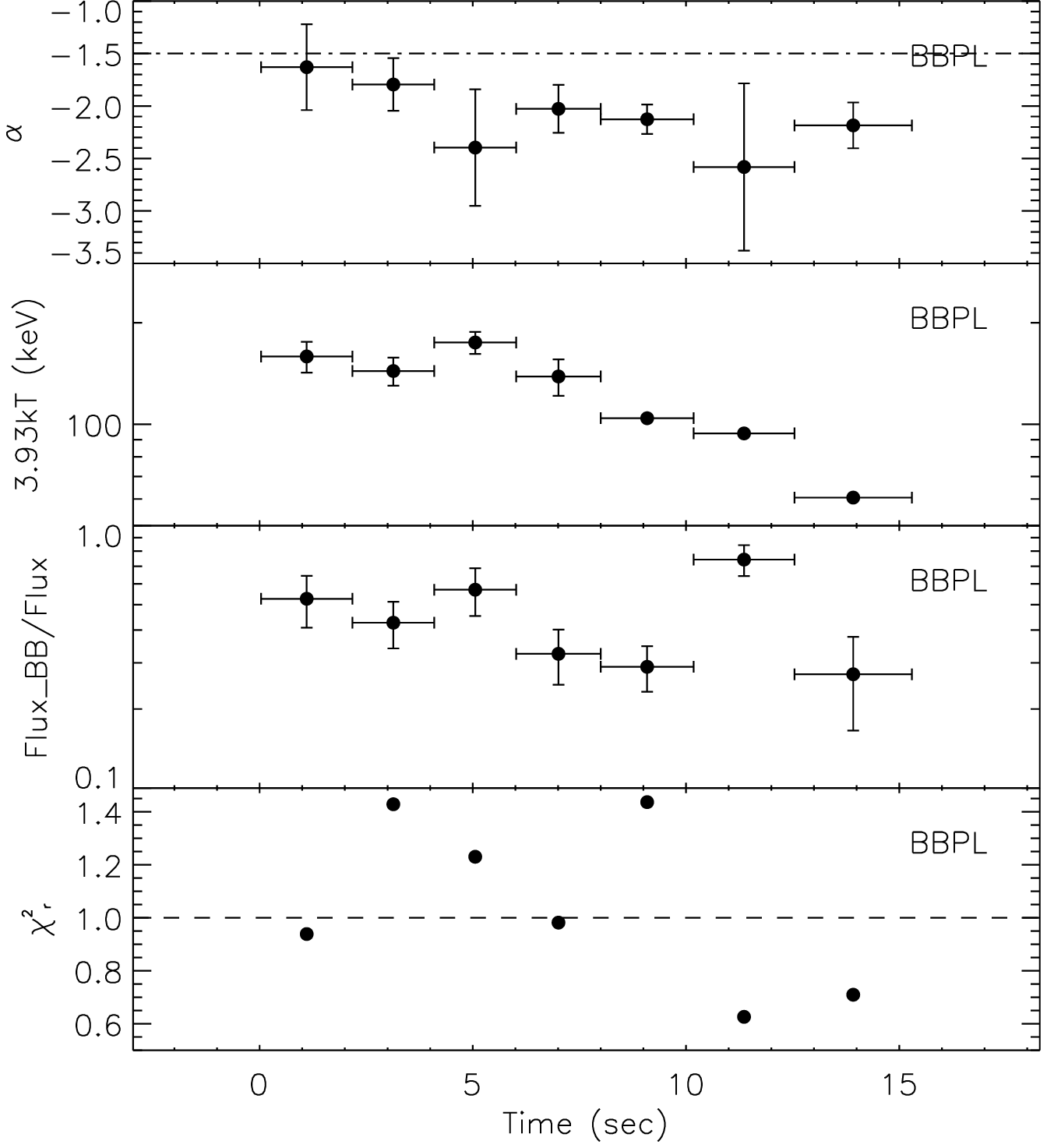,height=6.5cm,width=8.5cm}
\caption{Spectral evolution of GRB 980425. Symbols are the same of
  Fig.\ref{971214evo}.}
\label{980425evo}\end{center}
\end{figure}

GRB~971214 (\batse\ trigger 6533) has a highly variable light curve (Fig.
\ref{971214evo} -- top panels) and the time resolved spectral analysis could
be performed  on $\sim$ 20 s of the total GRB duration.  In this time
interval we extracted 13 spectra.  In Fig. \ref{971214evo} we show the time
evolution of the spectral parameters for the B, CPL and BBPL model 

The low energy spectral index $\alpha$ of the B and CPL model evolves
similarly, and for most spectra this parameter violates the optically thin
\syn\ limit (i.e. $\alpha=-2/3$; dashed line in Fig. \ref{971214evo}) and, of
course, the optically thin synchrotron limit in the case of radiative cooling
(i.e. $\alpha=-3/2$; dot--dashed line in Fig. \ref{971214evo}).  
In the case of the BBPL model instead $\alpha$ is 
always consistent with (i.e. softer than) these
limits and softer than the corresponding values found with the B or CPL model.
The peak energy of the three models is very similar and tracks the light curve
although it does not change dramatically.  
The BBPL fit shows that the peak energy of the black--body component tracks 
the light curve.  
The \bb\ can contribute up to $\sim$50\% of the total flux.

All three models give acceptable fits for the time integrated spectrum,
accumulated over 20 s.  The B model high energy component is very soft (i.e.
$\beta\sim 5$) making it consistent with the CPL model.  
For both these two models 
$\alpha\sim -0.66\pm0.08$ (1$\sigma$ uncertainty, see Tab. 2), 
consistent with the
value reported in Tab. 1 that was derived by fitting the WFC+GRBM \bsax\ 
data (Amati et al.  2002).

\subsubsection{GRB 980326}

For GRB~980326 (BATSE trigger 6660) both the duration and the light curve are
not available in the BATSE archive. By analysing the spectral evolution we
could extract only two spectra in approximately the total duration of the
burst ($\sim$5 sec)\footnote{This duration is consistent with the 9 s reported
  in Tab.1 by Amati et al. 2002, based on the \bsax\ observation.}.  The first
spectrum (from 0.09 to 1.56 s) is well fitted by the B and CPL models which
give similar results, i.e.  $\alpha=1.2\pm0.3$, $\beta_{B}=-3.4\pm0.7$ and
$E_{\rm peak}=52\pm27$ keV, with $\chi^2_r= 0.93$ (for 102 degrees of freedom)
and $\chi^2_r=0.94$ (for 103 degrees of freedom) for the B and the CPL model,
respectively.  The second spectrum (from 1.56 to 4.09 s) has $\alpha$ and
$E_{\rm peak}$ consistent with the first one.  These two spectra, fitted with
the BBPL model show a soft \pl\ component (i.e.  $\alpha_{BBPL}\sim -2.5$)
and peak energy of $\sim$ 74 keV (with $\chi^2_r= 1.0$).

The spectral parameters of the average spectrum of GRB~980326 are reported in
Tab. 2 and they are consistent with those reported in Tab. 1. 
The only difference is the sligthly larger value of
$E_{\rm peak, B}\sim65$ keV (with $\chi^2_r=1.02$) obtained here.

\subsubsection{GRB 980329}

GRB~980329 (BATSE trigger 6665) has a structured light curve (Fig. \ref{980329evo}
top panels) with at least two small peaks preceding two
major peaks of similar intensity. For the spectral evolution we could
accumulate 37 time resolved spectra within the $\sim$17 s duration
of the burst corresponding to its $T_{90}$. 

The low energy spectral index $\alpha$ evolves similarly in the B and CPL
model and its values are between the two synchrotron limits (i.e. --2/3 and
--3/2).  The fit with the BBPL model instead requires a very soft \pl\ 
component and a time evolution similar to that of the \pl\ index of both the B
and CPL model, but with a value which is always smaller than --3/2.

The peak energy seems to evolve differently in the B and CPL model.  In the B
model $E_{\rm peak}$ does not change much during the burst and remains below
$\sim$ 300 keV, whereas in the CPL model $E_{\rm peak}$ changes in time and
reaches $\sim$ 1 MeV in correspondence of the major peak of the light curve
(at 6 s).  The fit with the BBPL model instead presents a peak energy which
does not evolve much and, similarly to the B fit, stays constant at around 200
keV.  The \bb\ component contributes, at least, 40\% of the total flux (bottom
panel in Fig. \ref{980329evo}). 

The average spectrum of GRB~980329 has been accumulated over its $T_{90}$ and
fitted with the three models.  We found $\alpha=-0.93\pm0.1$,
$\beta=-2.4\pm0.2$ and $E_{\rm peak}=253\pm 10$ keV (Tab. 2) for the fit with
the B model.  These spectral parameters are in good agreement (except for a
softer low energy spectral index) with those found by the fitting of the
\bsax\ data by combining the WFC/GRBM data (Amati et al.  2002) reported in
Tab. 1.  We note that a few time resolved spectra of this burst and also the
time integrated spectrum have a quite large $\chi^2_r$ when fitted with all
the three models. We suspect that this is due to the fact that these spectra
are characterized by very small statistical errors.  Indeed, we found that the
use of a 2\% of systematic errors uniformly distributed in the spectral range
makes the fits acceptable. However, to the best of our knowledge, this has not
been treated in the published literature. For this reason we list the spectral
results as they were obtained without accounting for additional systematic
uncertainties.  When accounting for systematic errors, the $\chi^2$ improves,
the fitted parameters remain unchanged and their associated uncertainties
slightly increase.

\subsubsection{GRB 980425}

GRB~980425 (BATSE trigger 6707) is a long single peaked smooth GRB famous for
being the first GRB associated with a SN event (i.e. SN1998bw --
Galama et al. 1998).  GRB~980425 is also the lowest redshift GRB ever
detected.  Due to its relatively low fluence, its isotropic equivalent energy
is small compared to other bursts.  Indeed, it is one of the two 
clear outliers (the
other being GRB~031203) with respect to the \ama\ correlation (but see
Ghisellini et al.  2006).

To the aim of studying its spectral evolution we extracted 7 spectra during
roughly 15 s.  The time interval covered by our time resolved spectral
analysis is between the two durations $T_{90}$ and $T_{50}=9.79\pm0.29$ which,
however, differ by a factor 10.  This limitation is due to the slow decay of
the light curve after the trigger coupled to a relatively small intensity of
the burst.  As a result we could not constrain the spectral parameters of any
spectrum during the 15--33 s time interval.  However, our spectral analysis
covers the main part of the single pulse of the light curve and excludes only
the last decaying part of the light curve.

Although the B and CPL model can fit the time resolved spectra and give
consistent results (top and mid panels of Fig. \ref{980425evo}), we note 
that in 4 out of 7 spectra the B model yields an
unconstrained high energy spectral index $\beta$, suggesting that the CPL
model represents the data better.  The low energy spectral index $\alpha$ in
both cases is harder than the cooling limit, and for 3 out of 7 spectra it
also violates the optically thin synchrotron limit.  The evolution of the
peak energy is smooth and it decreases monotonically from $\sim$ 200 keV at
the beginning to few tens of keV in the final part of the burst.

The fit with a BBPL model (Fig. \ref{980425evo} bottom panel) gives a soft
\pl\ index, remaining softer than $-3/2$ during the burst evolution.
Overall we note that the \bb\ contribution to the total flux is around 40\%
except for one spectrum that has a quite considerable \bb\ flux (i.e.
$\sim$80\%). 
The peak energy (in this case the peak of the \bb\ component) is consistent, 
in terms of values and evolution, with that of the B and CPL model.

The time integrated spectrum, accumulated over the 33 s of duration of the
burst, is well fitted by the three models although, also in this case, the B
model has $\beta$ unconstrained.  The low energy spectral index of the time
integrated spectrum is $\alpha=-1.26\pm0.14$ and the peak energy is $E_{\rm
  peak}=123\pm36$ keV (Tab. 2), consistent with those reported in Tab. 1. The
BBPL model fitted to the time integrated spectrum gives a very soft \pl\ 
($\alpha=-2.19\pm0.16$) and a peak energy of the \bb\ component $E_{\rm
  peak}\sim 137$ keV, which is consistent with the fit obtained with the CPL
model.

\subsubsection{GRB 990123}

\begin{figure}
\begin{center}
\psfig{figure=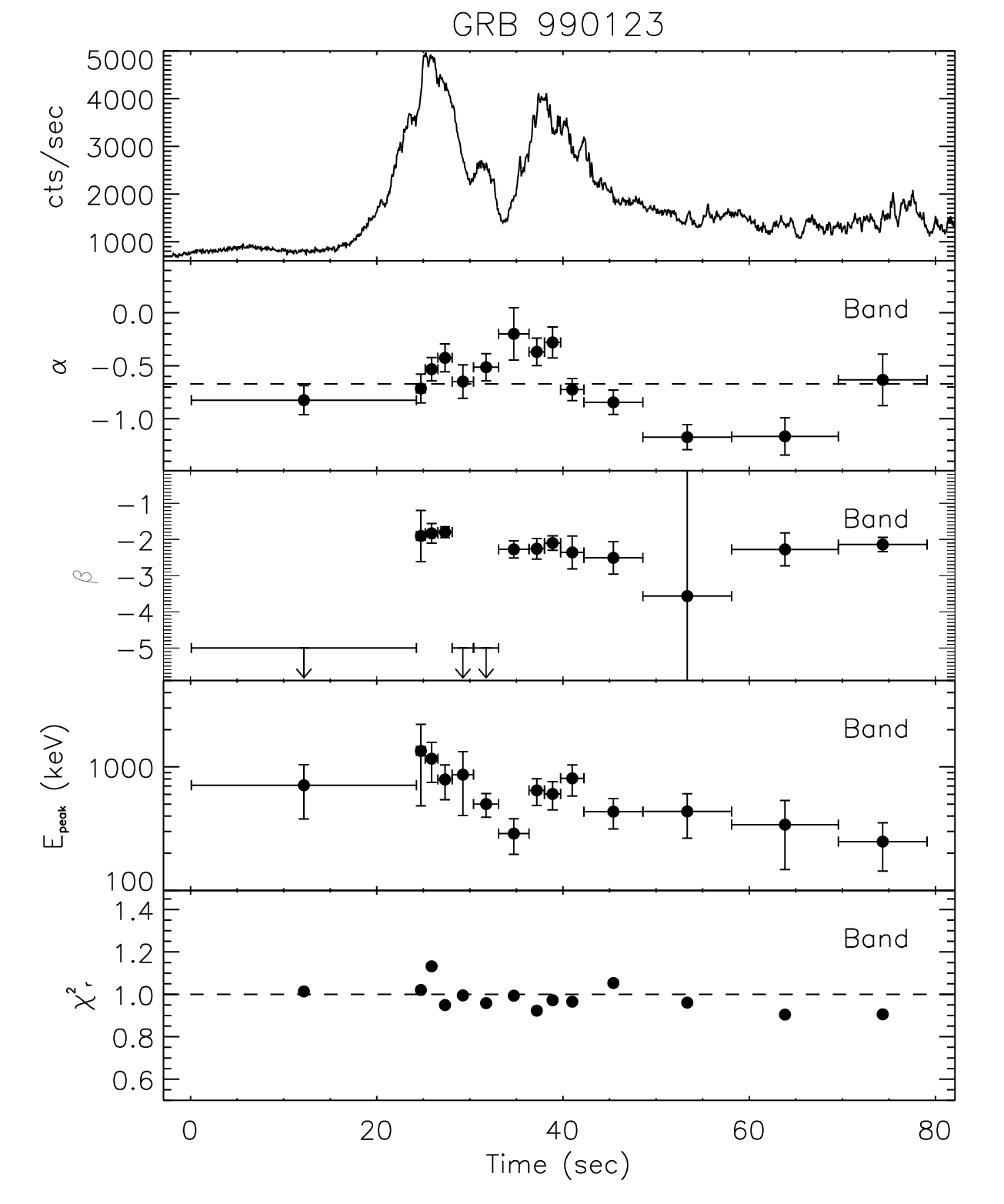,height=7.5cm,width=8.5cm}
\psfig{figure=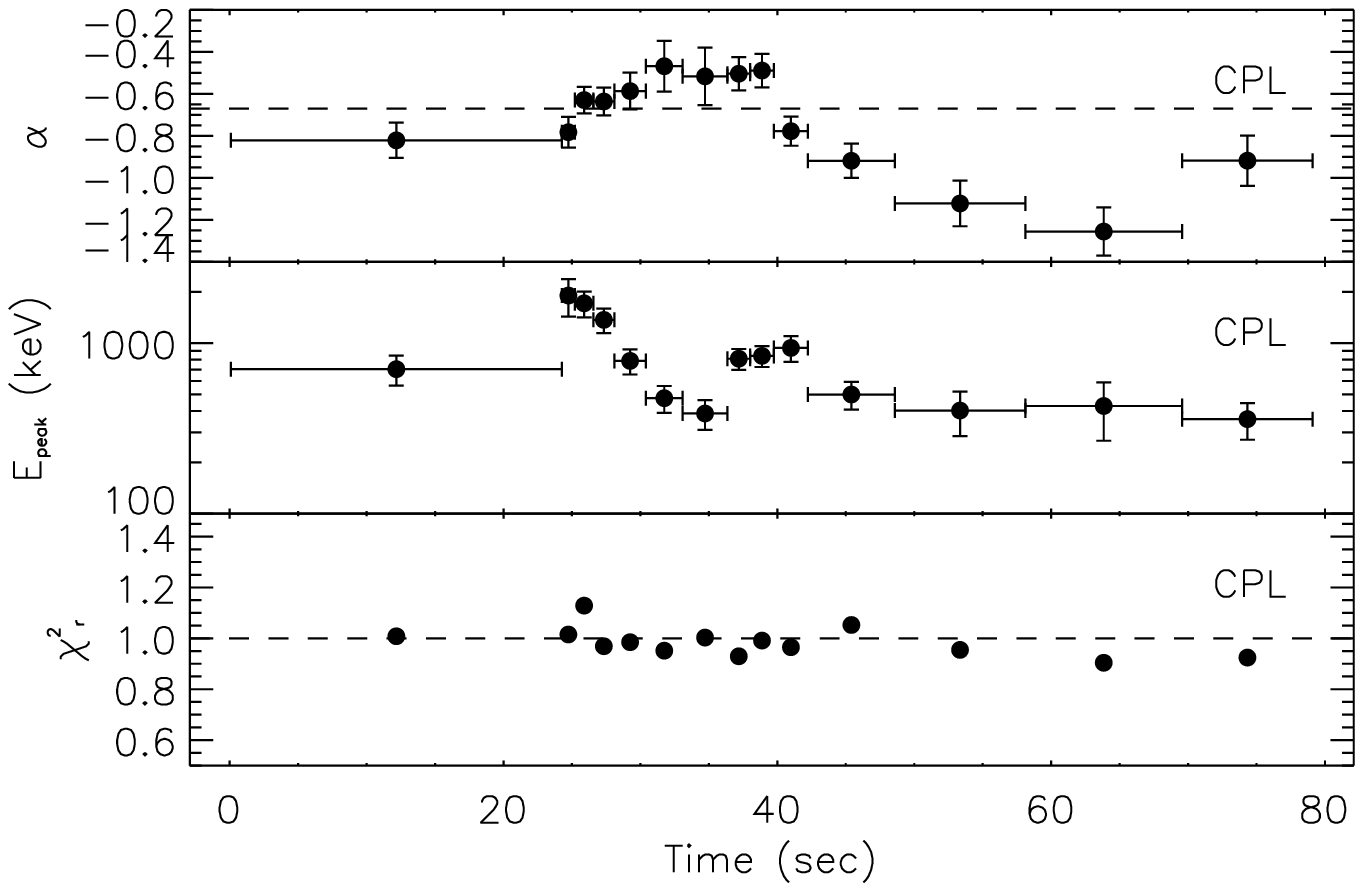,height=5cm,width=8.5cm}
\psfig{figure=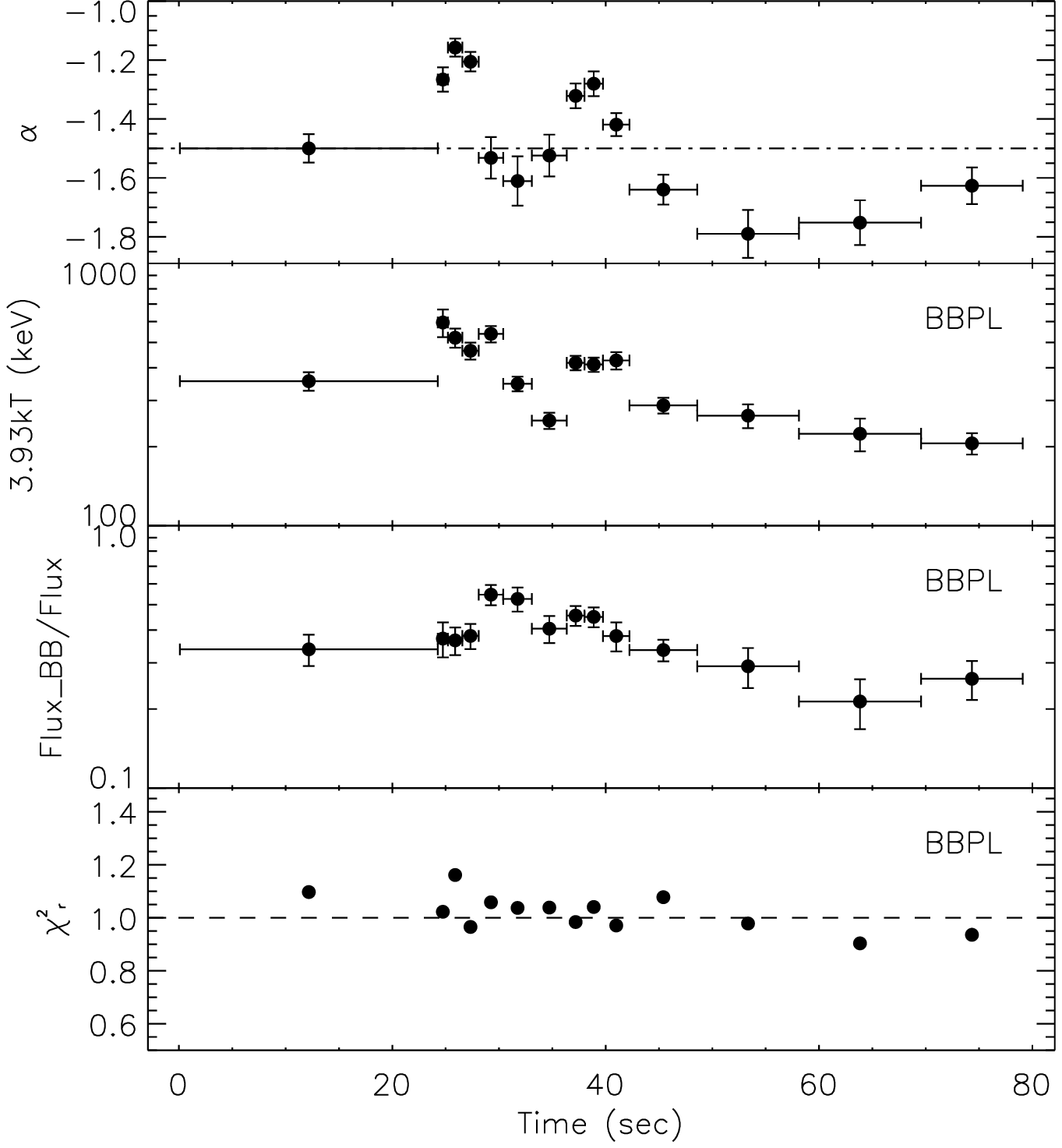,height=6.5cm,width=8.5cm}
\caption{Spectral evolution of GRB 990123. Symbols are the same of
  Fig.\ref{971214evo}.}
\label{990123evo}\end{center}
\end{figure}

GRB~990123 (BATSE trigger 7343) is a long duration event with a very high
fluence.  The light curve has two major peaks and a long tail after the second
peak.  There is a gap in the LAD data corresponding to the beginning of the
burst up to 20 s. For this reason we used the SD data.  The spectral evolution
(Fig.  \ref{990123evo}) shows that the peak energy slightly precedes the light
curve first peak while it tracks the second peak (see e.g. Ghirlanda, Celotti
\& Ghisellini 2002).  The low energy spectral component is harder than the
synchrotron limit during most of the two major peaks.  The B and CPL model
have similar time resolved spectral parameters.  The BBPL model fits the time
resolved spectra with a \pl\ component which is harder than the $-1.5$ limit.
The \bb\ flux is no more than 50\% of the total flux.

The time integrated spectrum accumulated over $\sim$100 s (in order to include
the long tail of the second peak) is fitted by both the B and the CPL model.
These models give similar results: the low energy spectral index is
$\alpha=-0.85\pm0.04$ (B) and $\alpha=-0.9\pm0.03$ (CPL); the peak energy is
$E_{\rm peak}\sim 605$ keV (B) and $E_{\rm peak}\sim 684$ keV (CPL).  The
latter values are lower than those reported in Tab. 1.  This is likely due to
the better energy coverage of the BATSE data (with respect to the GRBM
spectrum -- Amati et al. 2002): the extension of the energy range up to 1800
keV allows to better determine the value of $E_{\rm peak}$.

\subsubsection{GRB 990510}
\begin{figure}
\begin{center}
\psfig{figure=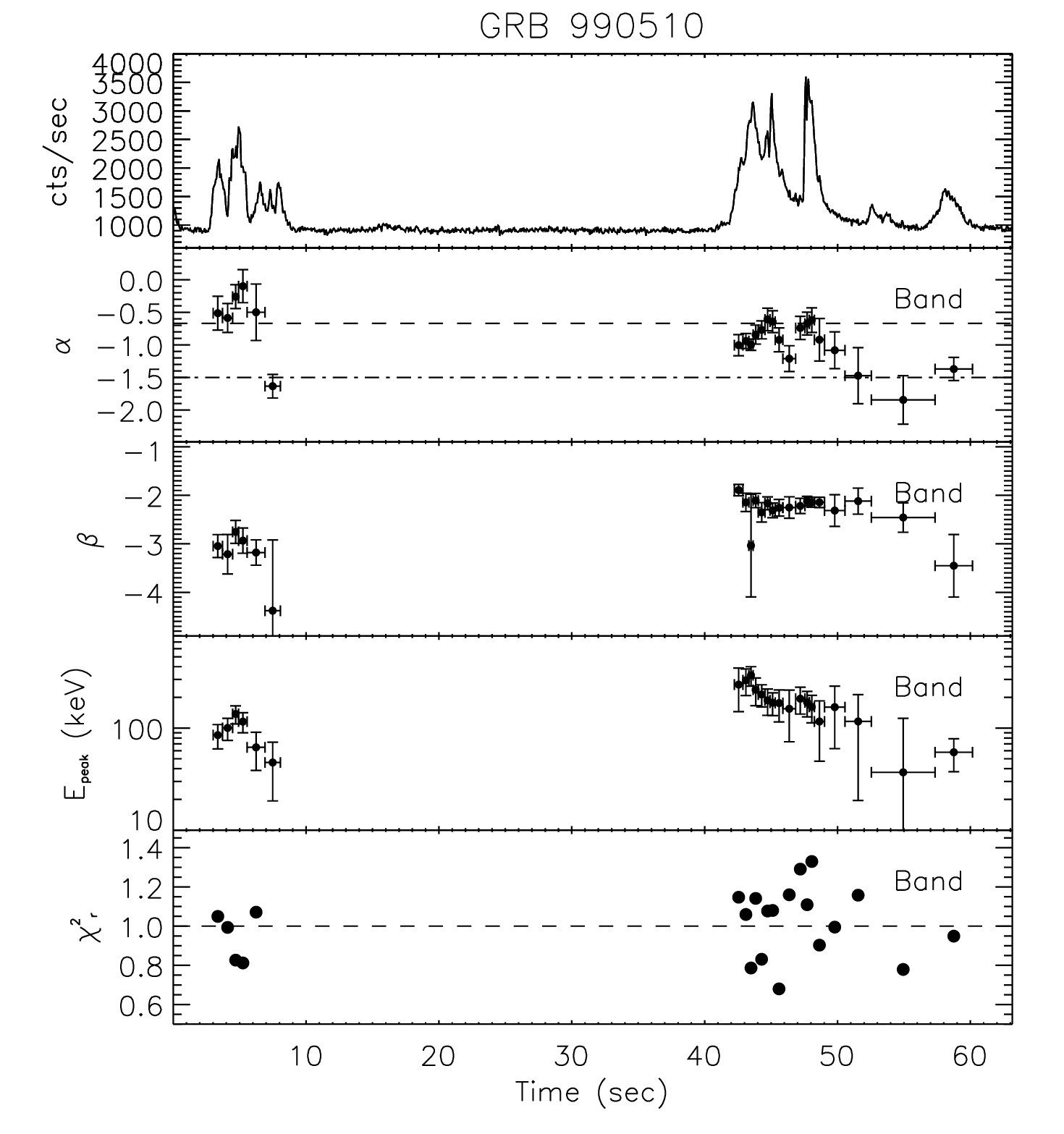,height=7.5cm,width=8.5cm}
\psfig{figure=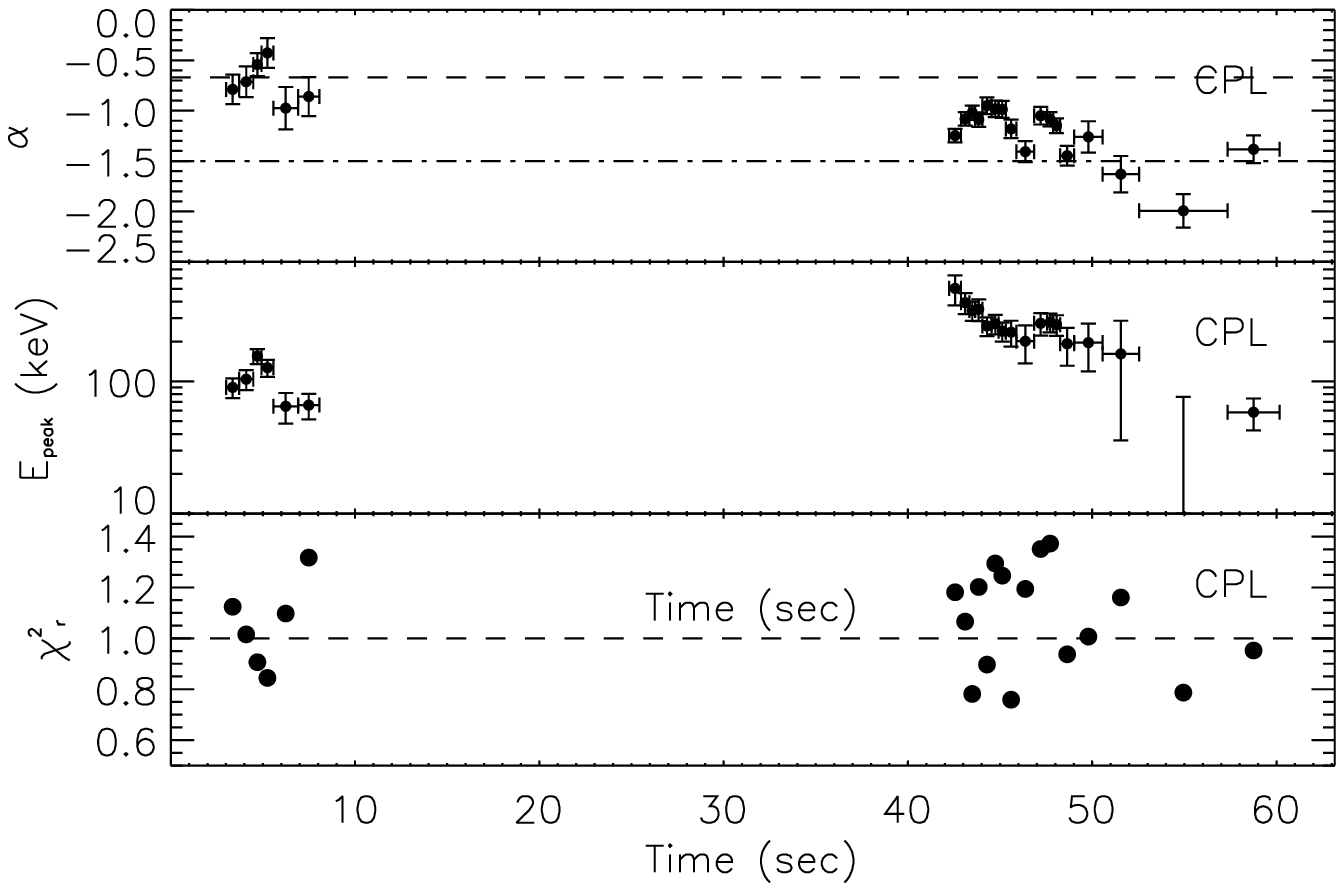,height=5cm,width=8.5cm}
\psfig{figure=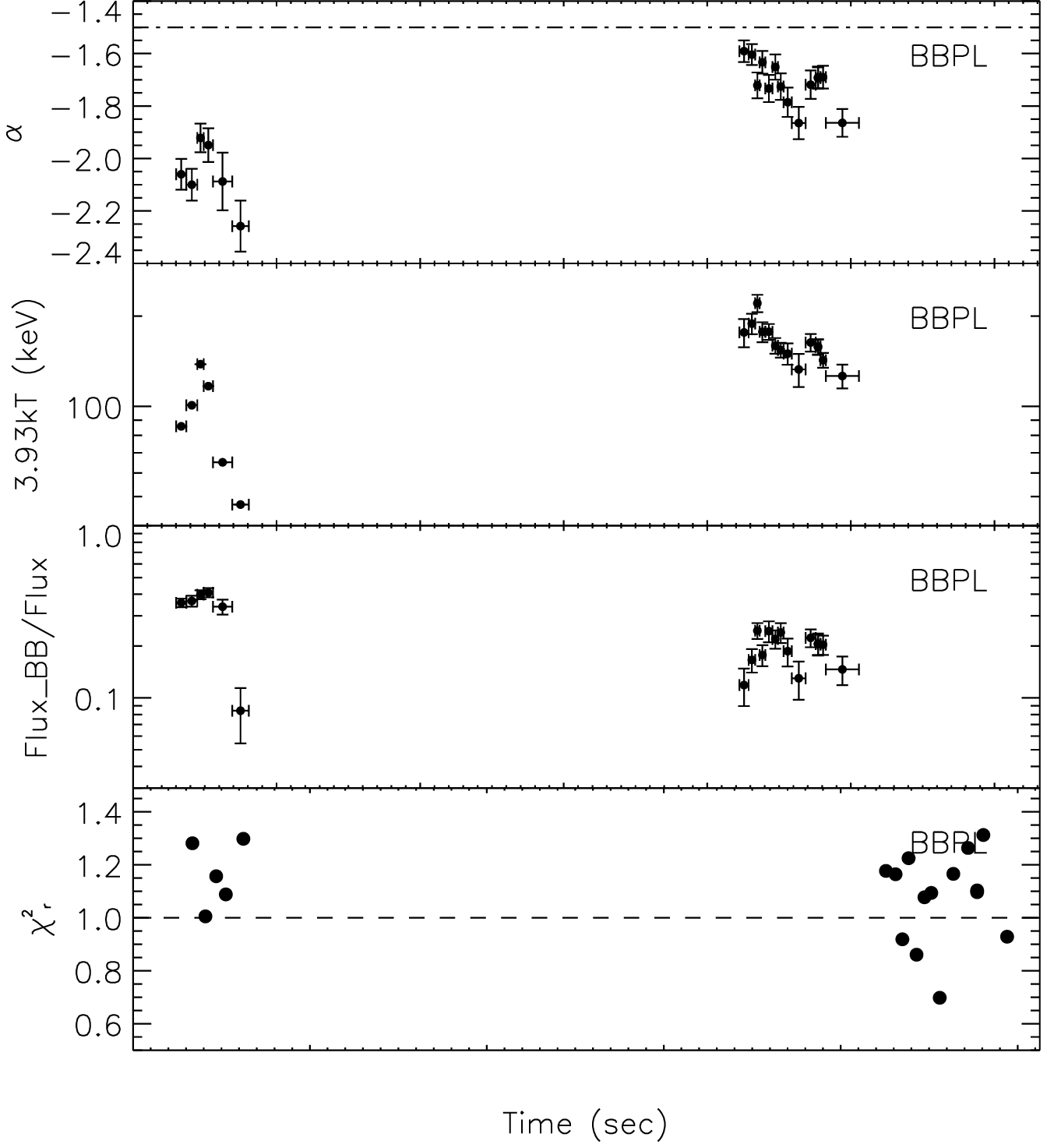,height=6.5cm,width=8.5cm}
\caption{
Spectral evolution of GRB 990510. Symbols are the same of
  Fig.\ref{971214evo}. }
\label{990510evo}\end{center}
\end{figure}

GRB~990510 (BATSE trigger 7560) has a light curve with two main structures
(lasting 10 and 20 s respectively) composed by several sub--peaks and
separated by a quiescent phase lasting $\approx$30 s. We could extract 6
spectra (distributed between 0 and 8 s) corresponding to the first set of
peaks and 17 spectra (between 40 and 60 s) corresponding to the second set of
peaks.  Given the long quiescent phase we analysed separately the time average
spectra integrated over the first and the second phase.

The time resolved spectra are well fitted with the CPL and the B model
which give similar results (see Fig. \ref{990510evo}).  The comparison
of the low energy spectral index and the peak energy between the first
and the second phase shows that the spectrum of the latter is (on
average) slightly softer in terms of $\alpha$ and harder in terms of
$E_{\rm peak}$ than the former. The low energy spectral index $\alpha$
is harder than the optically thin synchrotron limit for most of the
first peak and is consistent with this limit during the second
emission episode. $E_{\rm peak}$ rises and decays during the first
peaks while it has a more regular hard--to--soft evolution during the
second set of peaks.

The fit with the BBPL model (Fig. \ref{990510evo} bottom panels) is
consistent with the behaviour observed in previous bursts. In the case
of the first peak we could not constrain the \bb\ component of the
BBPL model. We therefore fixed, only for the time resolved spectra of
the first peak, the \bb\ temperature so that its peak corresponds to
the value found by fitting the B model. In the case of the BBPL model
the \pl\ component is much softer than the low energy component of the
CPL model and does not violate the $-3/2$ (cooling) limit. The peak
energy of the \bb\ component evolves similarly to that of the CPL (or
B) model and is slightly harder in the second emission phase than in
the first.  The \bb\ component contributes at most 30\% of the total
flux of the time resolved spectra.

The time integrated spectra of the first and second set of peaks have
been fitted separately (Tab. 2).
The spectral parameters of the fit of the second peak are consistent
with that reported in Tab. 1 obtained with the \bsax\ WFC+GRBM data
(Amati et al. 2002).

\begin{table*} 
\begin{center}
\begin{tabular}{llllllllll}
\hline
GRB    & Model  &$\alpha$        &$\beta$    &$E_{\rm peak}$&$\chi_{r}^{2}$& $\alpha_{PL}$  
&\% $F_{BB}$ & \% $F_{BB}^{BBCPL}$  \\
\hline
971214 &  CPL   &$-$0.65 (0.1)    & ....          & 186 (15) &  1.07        &$-$1.9  &36 &23    \\
980326 &  CPL   &$-$1.21 (0.44)   &  ...          & 65 (35)  &  1.02        &$-$2.7  &5  &$<$1  \\
980329 &  Band  &$-$0.93 (0.1)    &$-$2.4(0.1)    & 253 (10) &  1.6$^*$     &$-$1.7  &30 &26    \\
980425 &  CPL   &$-$1.26 (0.14)   &  ...          & 123 (36) &  1.04        &$-$2.1  &45 &8    \\
990123 &  Band  &$-$0.85 (0.04)   &$-$2.44 (0.23) & 607 (71) &  1.04        &$-$1.5  &38 &33    \\
990510 &  CPL   &$-$0.88(0.01)    &  ...          & 92 (6)   &  1.3$^*$     &$-$2.12 &32 &1.3   \\
       &  Band  &$-$1.16(0.05)    &$-$2.28(0.06)  & 173(21)  &  1.5$^*$     &$-$1.92 &18 &13     \\
\hline
\hline
\end{tabular}                                                                                           
\caption{ Time integrated properties of the bursts of our sample.
  Spectral parameters were obtained from the analysis of the time
  integrated spectrum of the \batse\ data. We report the best fit
  model parameters. For GRB~990510 we give the spectral results of the
  first and the second emission episodes separately. $*$ in these
  cases (see text) the reported $\chi^2_r$ (and the uncertainties
  associated to the spectral parameters) are without adding systematic
  errors to the fit (see text). $\alpha_{PL}$ represents the photon spectral
  index of the \pl\ component of the BBPL model fitted to the time integrated
  spectrum. $F_{BB}$ represents the average of the \bb\ contribution
  to the total flux obtained in the fits of the time resolved spectra.
  In the final column we show the contribution of the \bb\ component
  when fitting a more complicated model (see text) composed by a
  cutoff power--law plus a \bb.  In these fits the \bb\ peak energy
  has been fixed to the value obtained by the fit of a simple CPL
  model. These results represents an upper limit to the \bb\
  component, i.e. obtained by forcing the \bb\ to contribute to the
  peak of the spectrum. The reported \bb\ percentage is obtained by
  integrating in time the single contribution obtained by the fit of
  the time resolved spectra.  }
\end{center}
\label{tab2}
\end{table*}                                                                                                            

\subsection{Inconsistency of the black--body+power law model
with the Wide Field Camera data}
\begin{figure}
\begin{center}
\psfig{figure=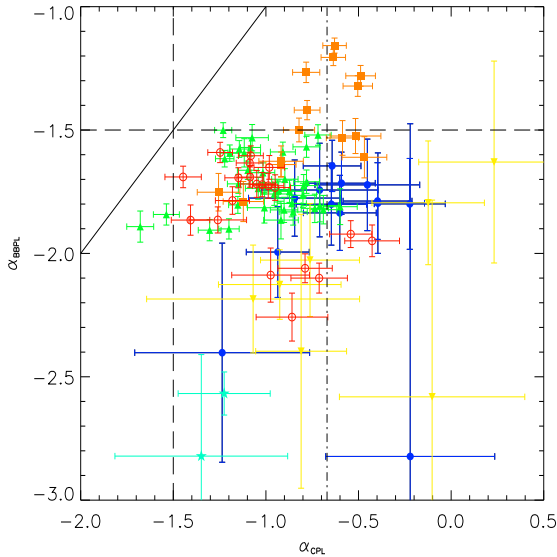,height=8cm,width=8cm}
\caption{
  Comparison of the photon index of the \pl\ component of the BBPL
  model ($\alpha_{\rm BBPL}$) with the low energy photon spectral index
  obtained from the CPL fit ($\alpha_{\rm CPL}$).  Different symbols
  correspond to: 971214 (filled circles), 980326 (filled stars), 980329
  (triangles), 980425 (upside--down triangles), 990123 (squares) and 990510
  (open circles). The solid line represents the equality of the two spectral
  indices.  The long--dashed line and the dot--dashed line are the synchrotron
  limits with and without cooling, respectively. }
\label{confalfa}\end{center}
\end{figure}

The results obtained from the time resolved analysis of the GRBs of our sample
indicate that the fit with a \bb+\pl\ model gives acceptable results for all
bursts.  This model has also the advantage, with respect to the Band and the
cutoff \pl\ model, to require a soft \pl\ component with a spectral index
always consistent (except for GRB 990123) with a cooling particle distribution
(i.e. $\alpha<-3/2$).  In Fig.  \ref{confalfa} we compare the photon index of
the CPL model (which is in most cases consistent with $\alpha$ of the B model)
to that the BBPL model.  Note that the latter is always softer than the
corresponding parameter of the CPL model.  In the same plot we also mark the
synchrotron limits and show that the \pl\ of the BBPL model is consistent
with these limits being (except for GRB 990123) softer than $-3/2$. Also when
considering the time integrated spectra we find that the \pl\ component of
the BBPL model is systematically softer than the \pl\ components of the B
or CPL model (compare col. 7 and col. 3 in Tab. 2).

The peak energy $E_{\rm peak}$ resulting from fitting the data with the BBPL
model is indeed produced by the \bb\ component which substantially contributes
to the total energetics, at least in the observed energy range of \batse.
This would thus favour the ``\bb\ interpretation" of the spectral--energy
correlation which we have summarised in Sec. 2.

However, these results are based on the spectral analysis of the \batse\ 
spectra only. Although covering two orders of magnitude in energy, these data
do not extend below 20 keV and above 2000 keV.  The low energy limit is
particularly relevant here, since for these bursts we do have the information
of the low (2--28 keV) energy emission from the WFC of \bsax.  We can then
compare the result of the BBPL model with the flux and spectrum observed by
the WFC.  Since the latter concerns the time integrated spectrum, we should
then either add the single time resolved spectra to construct the total flux
and spectrum for each burst, or use the result obtained fitting the \batse\ 
time integrated spectrum.  In both cases we have to extrapolate the model to
the energy range of the WFC.

\begin{figure}
\begin{center}
\psfig{figure=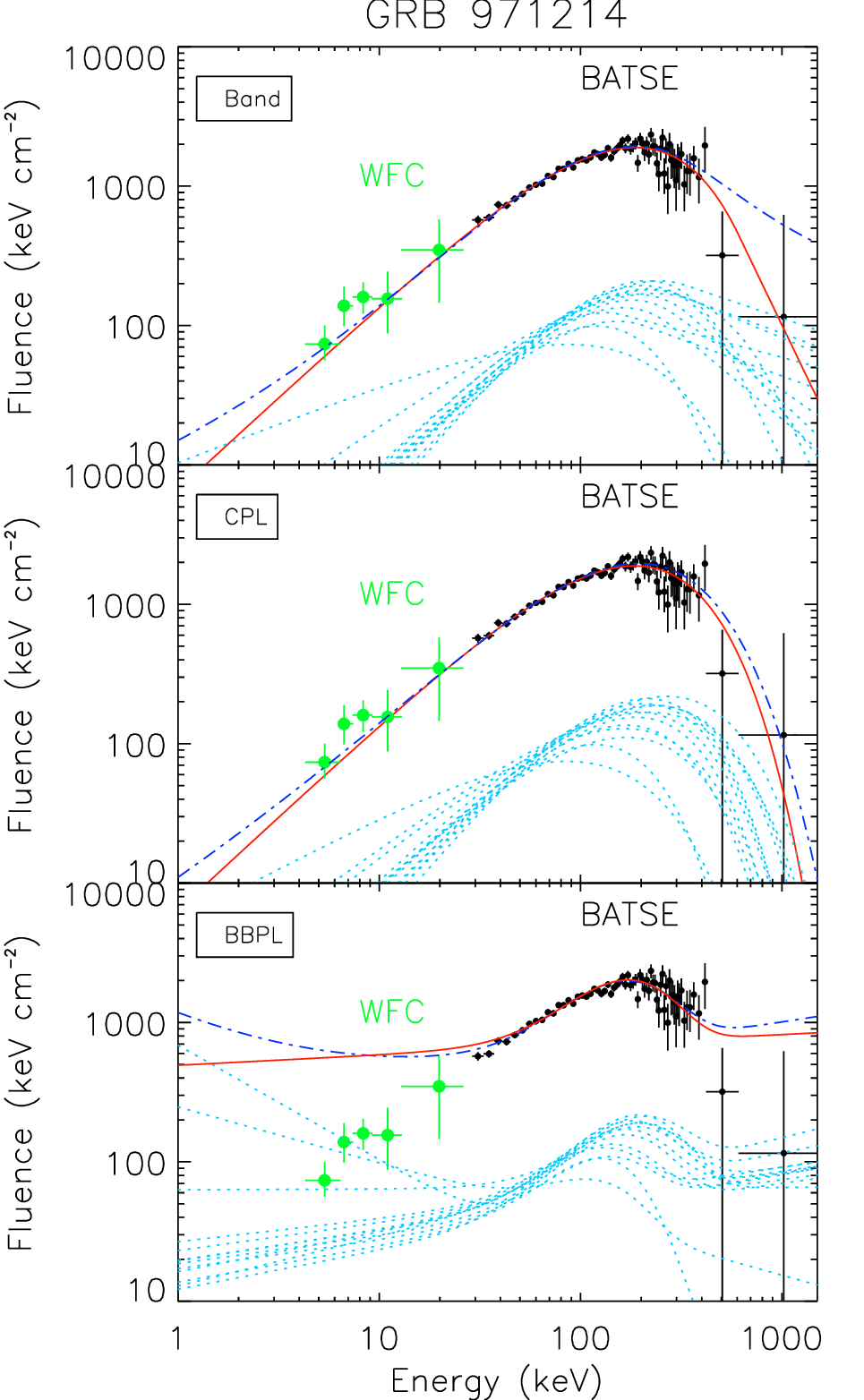,height=15cm,width=9cm}
\caption{ 
  GRB 971214: BATSE time integrated spectrum and WFC data (black and grey
  points, respectively).  In the three panels we show the spectral fits of the
  time resolved spectra (dotted lines), the spectral fit of the time
  integrated spectrum (solid line) and the sum of the time resolved spectral
  fits (dot--dashed line).  Spectral fits with the B model (top panel), CPL
  model (mid panel) and BBPL model (bottom panel) are show.  }
\label{971214wfc}
\end{center}
\end{figure}
\begin{figure}
\begin{center}
\psfig{figure=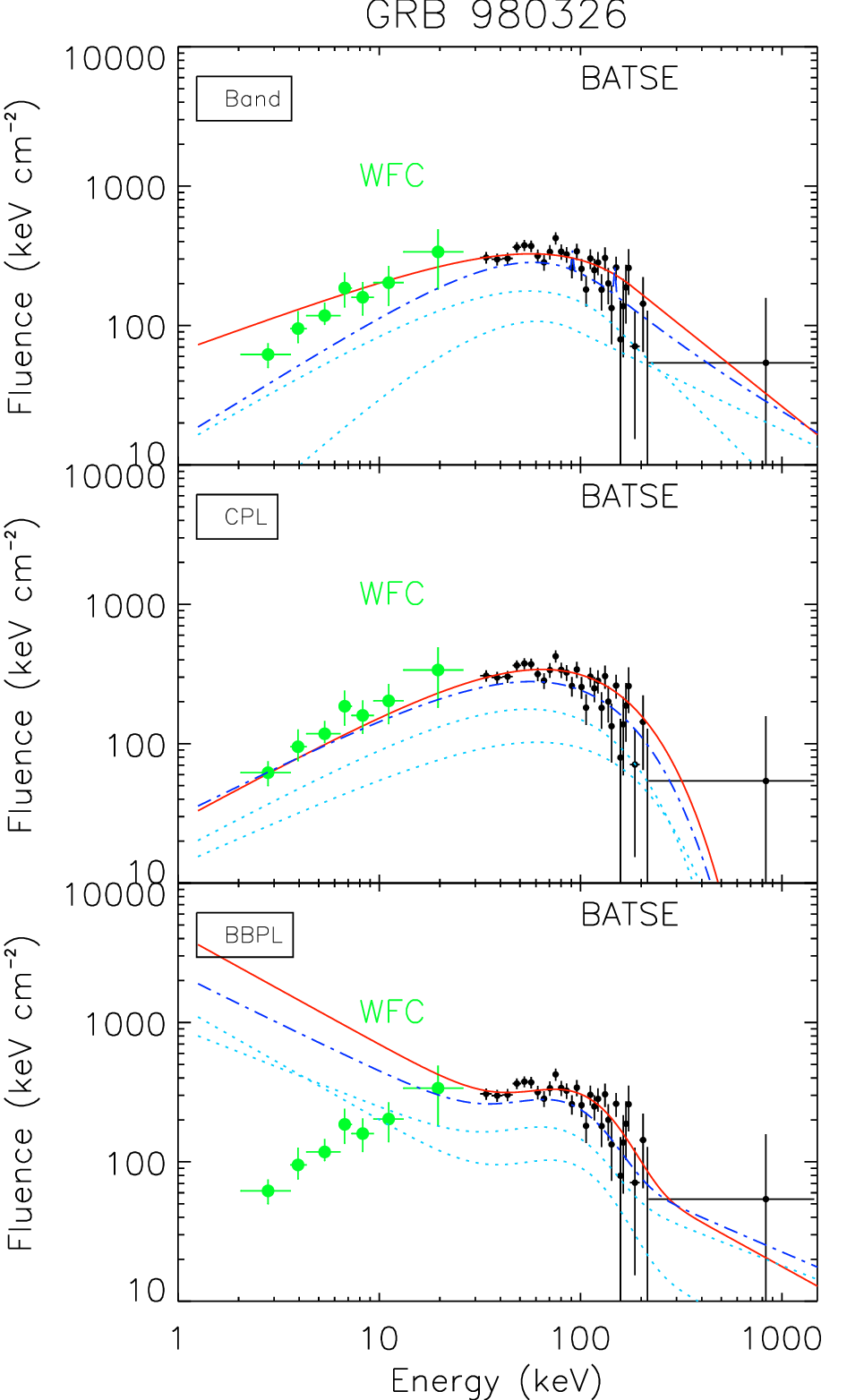,height=15cm,width=9cm}
\caption{
  GRB 980326. Symbols are the same of Fig.\ref{971214wfc} }
\label{980326wfc}
\end{center}
\end{figure}
\begin{figure}
\begin{center}
\psfig{figure=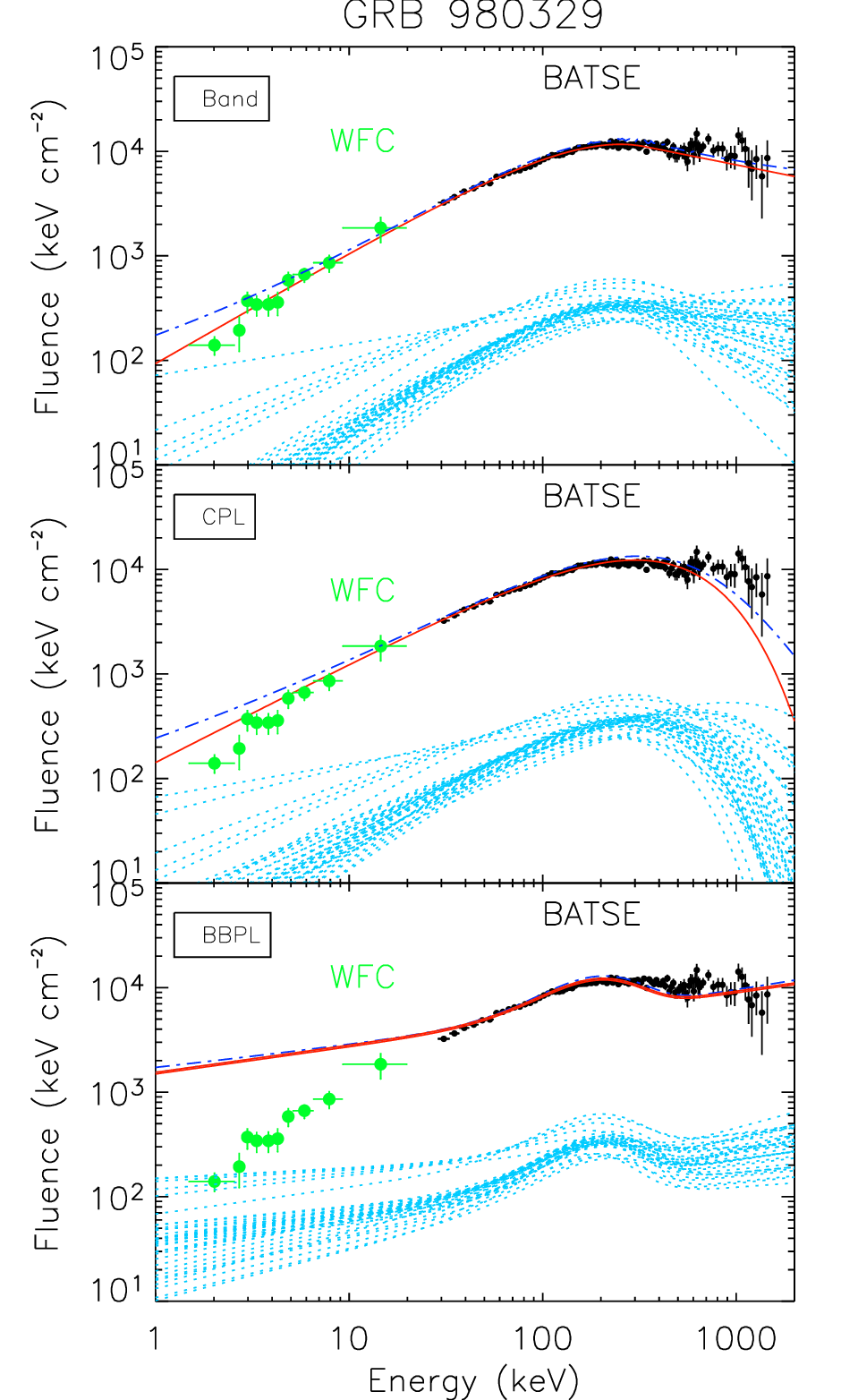,height=15cm,width=9cm}
\caption{
GRB 980329. Symbols are the same of Fig.\ref{971214wfc} }
\label{980329wfc}
\end{center}
\end{figure}
\begin{figure}
\begin{center}
\psfig{figure=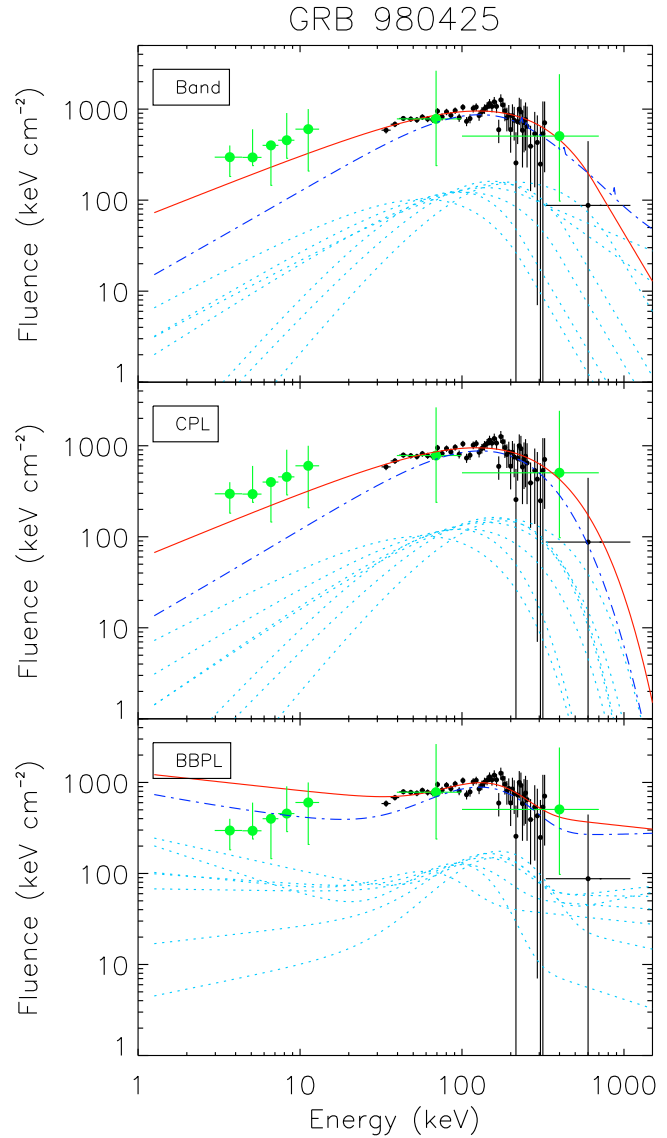,height=15cm,width=9cm}
\caption{
  GRB 980425.  Symbols are the same of Fig.\ref{971214wfc}. In this case we
  also show the two data points of the GRBM instrument on board \bsax\ 
  covering the 40-700 keV energy range.  }
\label{980425wfc}
\end{center}
\end{figure}
\begin{figure}
\begin{center}
\psfig{figure=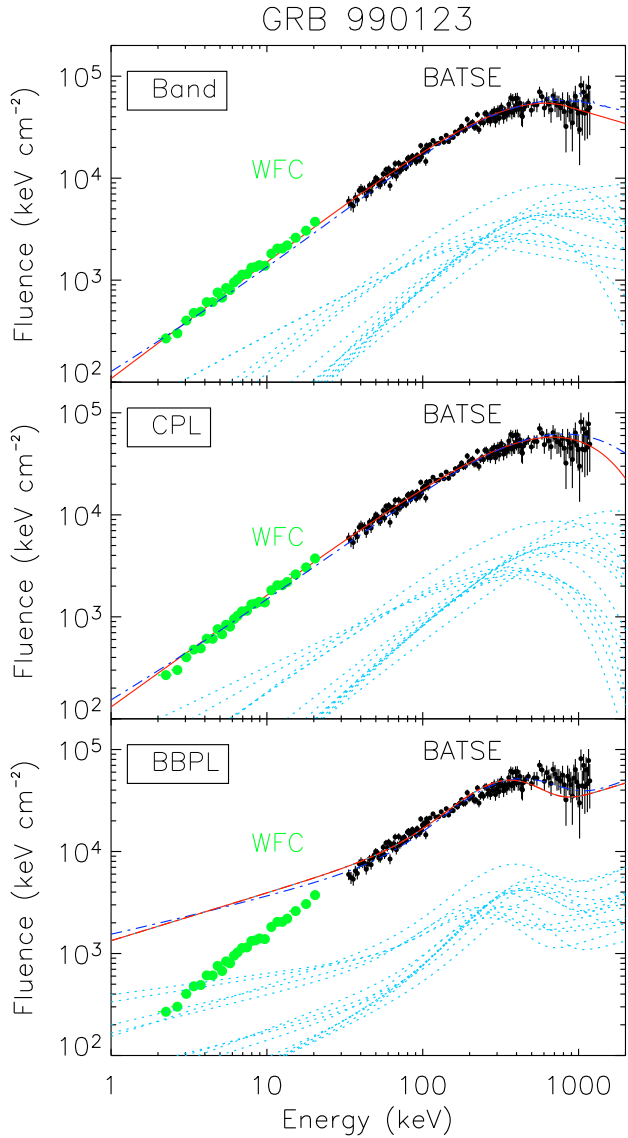,height=15cm,width=9cm}
\caption{
  GRB 990123. Symbols are the same of Fig.\ref{971214wfc} }
\label{990123wfc}
\end{center}
\end{figure}
\begin{figure}
\begin{center}
\psfig{figure=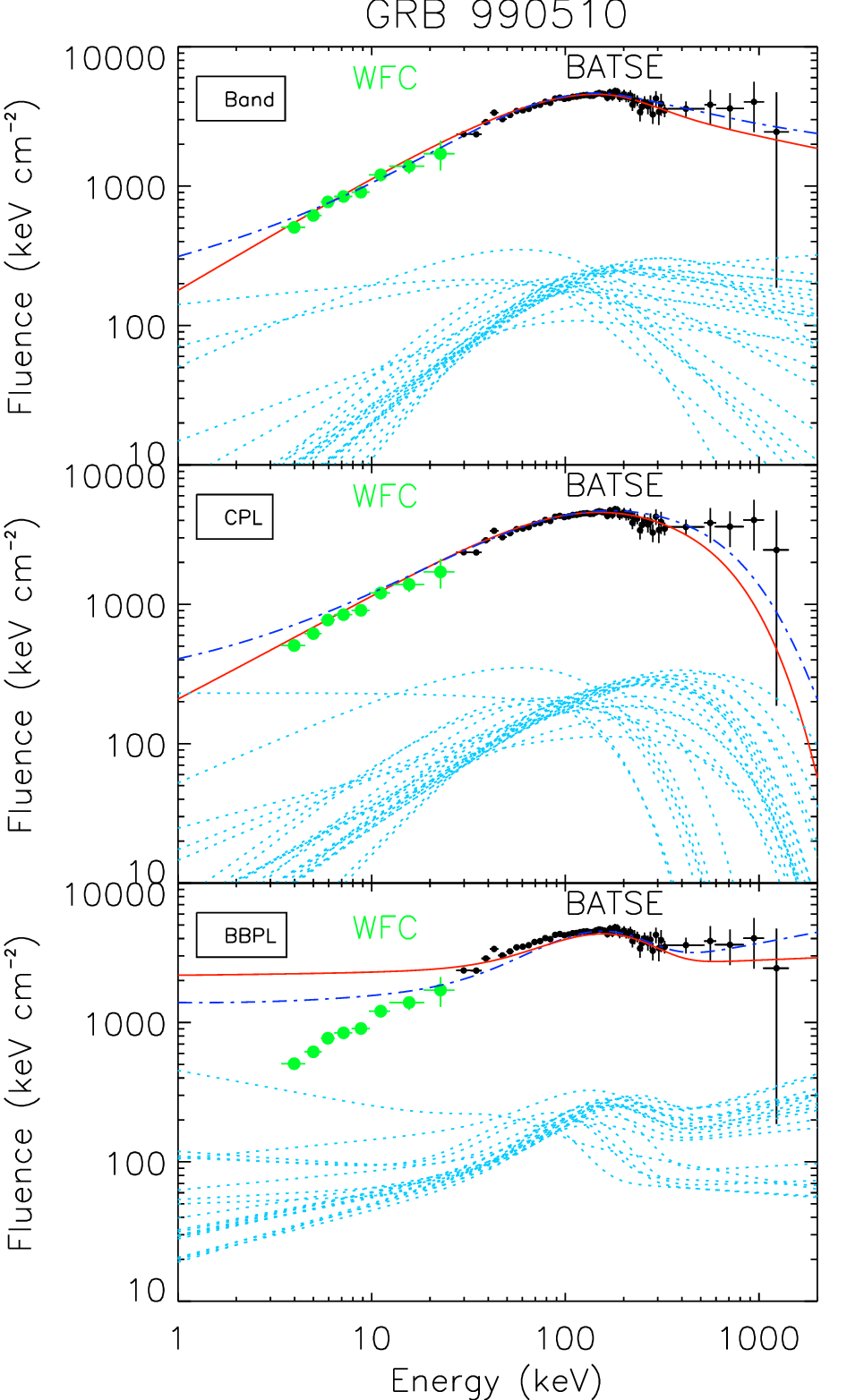,height=15cm,width=8.8cm}
\caption{
GRB 990510. Symbols are the same of Fig.\ref{971214wfc} }
\label{990510wfc}
\end{center}
\end{figure}

As stated above, the inclusion of the \bb\ component implies that the
accompanying \pl\ component becomes soft (i.e. $\alpha <- 1.5$).  It is this
\pl\ component that mainly contributes at low energies, and we find, in all
cases, a strong disagreement between the extrapolated flux and spectrum of
the WFC data.

This is shown in Figg.  \ref{971214wfc}, \ref{980326wfc}, \ref{980329wfc},
\ref{980425wfc}, \ref{990123wfc}, \ref{990510wfc}, where we report the BATSE
time integrated spectrum and the WFC spectrum.  In the three panels of these
figures we report the results of the fit with the three models described in
Sec. 3, i.e. the Band model (B), the cutoff \pl\ model (CPL) and the composite
model (black--body plus \pl\ -- BBPL). We report both the model fit to the
time integrated spectrum (solid line) to the time resolved spectra (dotted
lines) and the sum of the time resolved model fits (dot--dashed line).

One can see that in all cases the BBPL model strongly overpredicts the
observed flux in the WFC 2--28 keV energy band, with a slope which is much
softer than observed.  This occurs both when we sum the time resolved spectra
and when we use the time integrated fits.  On the contrary, note the excellent
agreement of the extrapolated flux and the WFC data in the case of the B
and the CPL fits.  To the best of our knowledge, this is the first time that a
detailed comparison of the WFC $Beppo$SAX and the \batse\ data is performed.
We conclude that they are in excellent agreement {\it if the spectrum is
  indeed described by the Band or CPL model, and that the BBPL model cannot
  reproduce the WFC data.}

We can also conclude that a fit with a \bb\ only (without the \pl)
is never consistent
with the data, even when considering spectra at the peak of the light
curve or for the first phases of the emission.
This is because fitting the CPL model, which can mimic a \bb\ when
$\alpha=1$, always gives $\alpha<0$.

Our analysis also shows that the black--body component in the time
resolved spectra that we have analyzed (typically with $>$0.1 s
time resolution) does not change much during the burst. This implies that
even if it were possible to perform the spectral analysis with a finer
temporal resolution, it is unlikely  that the time resolved
spectra are the superposition of a multi--temperature black--body.

Finally, we cannot exclude the possibility that the
{\it istantaneous} spectrum is produced by a superposition
of black--body components.
Indeed, this is exactly what happens in thermal or
quasi--thermal Comptonization models (if the seed 
photons have a relatively narrow range of frequencies),
where the superposition of different scattering orders
(each one being black--body like) produces
the cut--off power law spectrum.
Black--body components produced in different (and independent)
emitting regions, instead, are less likely, since some fine tuning
is required in order to produce the smooth observed spectrum.

\subsubsection{Further testing the black--body component}

The existence and the relevance of a \bb\ component in the spectra of
our GRBs can be further tested allowing for the possibility that the
real spectral model is more complicated than what we thought.  We
could make the \bb+\pl\ model fits consistent with the WFC [2--28 keV]
spectra by introducing a spectral break between the BATSE and the WFC
energy ranges.  This could indeed be the case if the non--thermal
component is produced by an electron energy distribution with a low
energy cutoff, or if the apparently non--thermal component is instead
the result of a thermal Comptonization process (e.g.  Liang 1997;
Liang et al. 1997; Ghisellini \& Celotti 1999; Meszaros \& Rees 2000).
In the latter case what we see in the WFC could be the (hard) spectrum
of the seed photons, while in BATSE we may see the sum of the
Comptonization spectrum and a \bb.

We must then check if, in this case, it is possible that a \bb\ is present, 
is responsible for a significant fraction of the total flux and for the 
observed $E_{\rm peak}$, without violating any observational constraint.
If so, then the ``\bb'' interpretation presented in \S 2 would
receive support.

However, there are severe problems with this possibility.
The first is that the required break should always be at $\sim$30 keV
(between the BATSE and the WFC energy ranges) despite the fact
that our GRBs have different redshifts. 
This makes this possibility rather ad hoc.

The second problem comes from the following test.
As stated, we should use a model composed by \bb\ plus a Band
spectrum. 
This model, unfortunately, has too many free parameters to
yield strong constraints, but we can mimic it by adopting
a model composed by the sum of a \bb\ and cutoff \pl.
The index of the latter should be thought as 
the low energy index of the Band model.
Furthermore, since what we really put on test is the
presence of a relevant \bb, we can also fix its temperature
requiring it to give the $E_{\rm peak}$ found when using
the CPL (or B) model. 
This is because we already know that are these $E_{\rm peak}$,
when combined in the time integrated spectrum, that give the
$E_{\rm peak}$ used for the Amati and Ghirlanda correlations.

We thus use this \bb+cutoff \pl\ model (BBCPL):
\begin{equation}
N(E)=A \frac{E^2}{\exp(E/kT)-1} + B E^{\alpha} \exp\left(-\frac{E}{E_{0}}\right)
\end{equation} 
where $kT$, i.e the black body characteristic temperature, is fixed so that
3.9$kT$=$E_{\rm peak}$ (as found from the fit of the CPL model to each time
resolved spectrum). 
This model has the same number of free parameters of the BBPL and B model
(the two normalisations, $E_0$ and $\alpha$).

\begin{figure}
\begin{center}
\psfig{figure=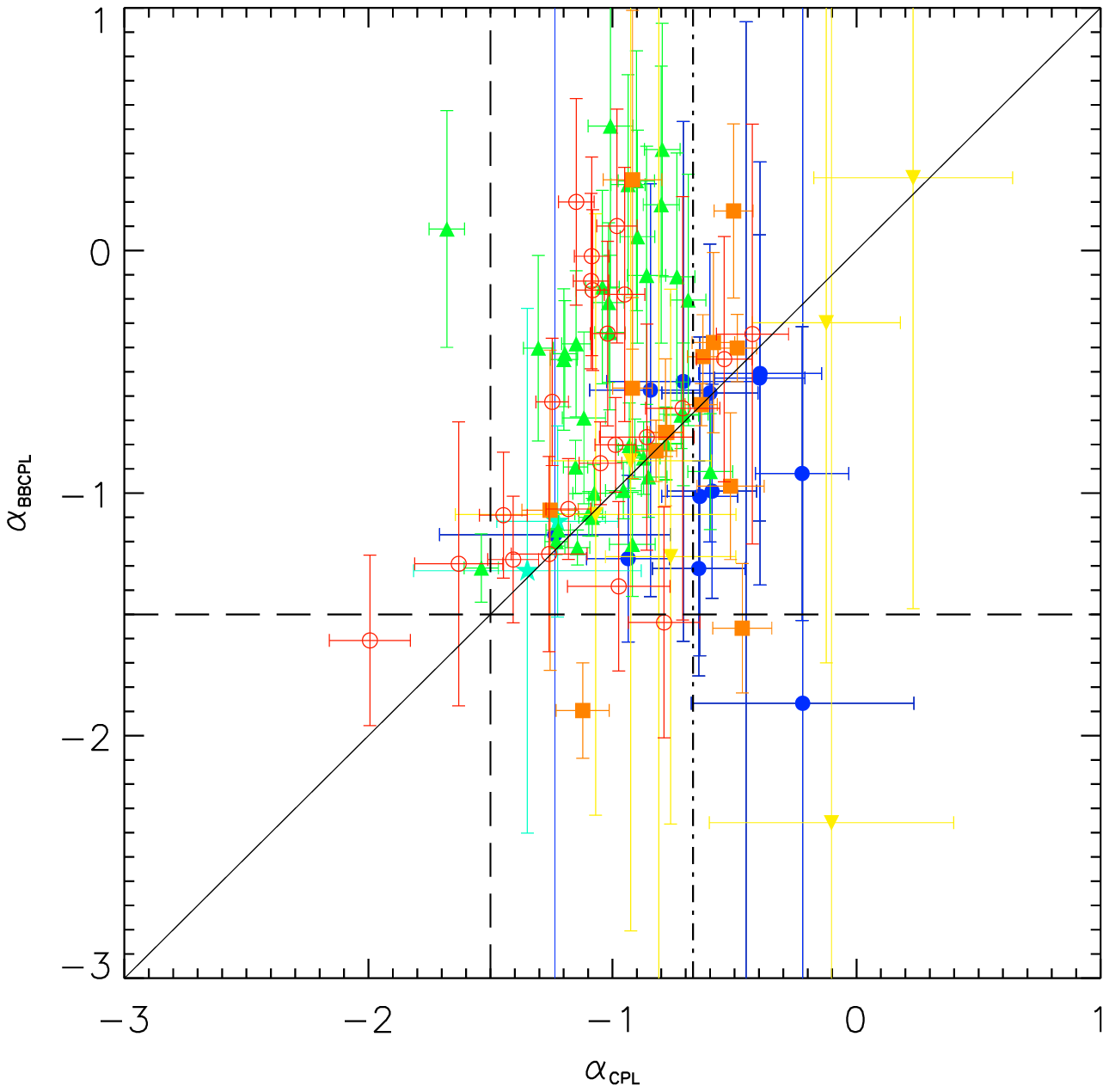,height=8cm,width=8cm}
\caption{
  Comparison of the spectral photon index of the CPL model
  ($\alpha_{\rm CPL}$) with the photon index obtained from the fit of
  a cutoff \pl\ plus a \bb\ with the peak of the \bb\ fixed to
  the values found from the fit of a simple CPL model. Symbols are as
  in Fig. \ref{confalfa}}
\label{confalfa1}\end{center}
\end{figure}

In Fig. \ref{confalfa1} we compare the photon index found with a simple CPL
model and the $\alpha$ of the BBCPL model described above. In the BBCPL model
the photon index of the CPL component can fit the WFC data and
indeed we found it to be consistent with the values found by the fit of a
simple CPL model. 
{\it Instead, the \bb\ component is negligible in all these fits.
}

For each time resolved spectrum fitted with the BBCPL model we can
compute the fraction of the rest frame bolometric flux contributed by the \bb\ 
component.  Summing up these contributions for the entire duration of each burst 
we derive the contribution of the \bb\ to the time integrated flux.
The values are reported in Tab. 2 (last column): for all the bursts this 
contribution is small.

We can then conclude that if a \bb\ is present, with a temperature consistent
with the peak of the spectrum (found by fitting the CPL model) then its flux
is not relevant. 
Consider also that this spectral model is not required by the data,
which are instead well described by the simpler CPL (or B) model.
In this sense what we found is an {\it upper limit} to
the possible contribution of a \bb\ to the total flux.

\section{Summary of results}

We have analysed the spectra of 7 GRBs detected by BATSE with measured
redshift and for which also the \bsax\ WFC spectrum has been published (Amati
et al. 2002).  We analysed both the time resolved and the time integrated
spectrum with three models: the Band model (B), a cutoff \pl\ model (CPL) and
a \bb+\pl\ model (BBPL).  For a further test of the importance of a possible
\bb\ component we have also used the sum of a \bb\ plus a cutoff \pl\ model
(BBCPL).  The comparison of the spectral parameters and the analysis of the
spectral evolution has shown that:
\begin{itemize}
\item the time resolved spectra could be reasonably fitted with all models.
  The spectral parameters of the B and CPL model agree within their
  uncertainties;
\item in all our GRBs the spectral slope of the low energy component of the B
  or CPL model violate both the optically thin synchrotron limit with ($-1.5$)
  or without ($-0.67$) radiative cooling;
\item the values of $\alpha<0$ found from the fit of the CPL model exclude the
  possibility that a single \bb\ model can fit these spectra (as the \bb\ 
  coincides with the CPL model only for $\alpha=1$);
\item the \pl\ slope of the BBPL model is softer than the corresponding
  parameter of the B or CPL model.  In most GRBs (except GRB 990123) this
  component is softer than the optically thin \syn\ limit with cooling
  ($-1.5$) and softens as time goes by;
\item the peak energies of the \bb\ component of the BBPL model found here are
  similar to the values found for a few other bursts analysed with the BBPL
  model (Ryde et al. 2005) or with a single \bb\ component (GCG03);
\item the \bb\ flux (in the BBPL model) is no more than 50\% of the total flux
  and it changes with time. In these bursts the \bb\ 
  does not dominate the initial emission phase as was the case of the few GRBs
  analysed by GCG03;
\item the soft \pl\ spectra found using the BBPL model implies a relatively
  large flux of the spectrum extrapolated at lower energies.  
  This extrapolation is inconsistent with the WFC data and spectra (Figg.
  \ref{971214wfc}, \ref{980326wfc}, \ref{980329wfc}, \ref{980425wfc},
  \ref{990123wfc}, \ref{990510wfc});
\item the time integrated spectral fit and the sum of the time
  resolved spectral fits with either the B and CPL model
  are consistent with the WFC spectrum both in terms of flux and slope;
\item fitting the \batse\ spectra with the BBCPL model results in a
  cutoff \pl\ component whose extrapolation to the WFC energy range is
  consistent with the observed spectrum in terms of flux and slope.  In this
  case, however, the \bb\ flux is not significant.
\end{itemize}

\section{Conclusions}

The most important results of this work is the assessment of the importance of
a \bb\ component in the spectra of GRBs.  For the GRBs analysed here, we find
that it cannot be, at the same time, responsible for the peak (in $\nu F_\nu$)
of the spectrum and for its total energetics.  We could reach this conclusion
by analysing the time resolved spectra of those GRBs detected by \batse\ and
the by the WFC of \bsax, therefore using the energy range between 2 keV and 2
MeV.  We also find that the \batse\ data, fitted by a cutoff \pl\ or by the
Band model, are entirely consistent with the WFC data. 

These findings bear important consequences on the interpretation of the
peak--energy correlations (including the Amati, the Ghirlanda, and the Firmani
correlations) put forward recently by Thompson (2006) and by Thompson,
Meszaros \& Rees (2007).  This interpretation requires that the \bb\ component
is responsible for the peak energy $E_{\rm peak}$ and for a significant
fraction of the bolometric emitted energy.  Note that, since the temperature of
the \bb\ component may vary in time, the time integrated spectrum may not be
particularly revealing of the \bb\ presence, making a time resolved analysis
mandatory.

One may argue that the spectrum is even more complex than what we thought,
having an additional break and becoming harder at low energies. 
Such a break is expected if the spectrum is due to a thermal 
photospheric emission (the \bb\ component) superimposed to non--thermal 
emission due to some dissipative mechanism (Meszaros \& Rees 2000). 
An alternative possibility is that the observed spectra result from 
multiple Compton up--scattering of soft seed photons 
(e.g. Ghiselllini \& Celotti 1999; Thompson 2005).
In such a case a break is expected between the (possibly) hard
seed photon spectrum and the beginning of the Comptonized spectrum.

But even by fitting the spectra with a more complex model 
allowing for this break, we found that the \bb\ component is not relevant.
This, together with the rather ad hoc requirement to have a break
always at 30 keV (observed) irrespective of the different redshifts
of our GRBs, lead us to conclude that the presence of relevant \bb\ 
in the spectrum of our GRBs is to be excluded.
This in turn, makes more problematic (and mysterious) the
interpretation of the spectral--energy correlation in long GRBs.

\section*{Acknowledgements}
We thank Annalisa Celotti for discussion and a PRIN--INAF 2005
grant for funding. We are grateful to the referee for her/his
comments. This research has made use of data obtained through the High
Energy Astrophysics Science Archive Research Center Online Service,
provided by the NASA/Goddard Space Flight Center.

\end{document}